\documentclass[apj]{emulateapj}  
\usepackage{graphicx}
\usepackage{amsmath}
\usepackage{comment}
\usepackage{txfonts}
\usepackage{natbib}			

\usepackage[percent]{overpic} 

\renewcommand\email\texttt

\begin{document}
 
\slugcomment{\sc submitted to \it Astrophysical Journal}
\shorttitle{\sc Distances to the Satellites of M31}
\shortauthors{Conn et al.}
 
\title{A Bayesian Approach to Locating the Red Giant Branch Tip Magnitude (Part II); \\
Distances to the Satellites of M31}

\author{A.\ R. Conn\altaffilmark{1, 2, 3}}

\author{R.\ A. Ibata\altaffilmark{3}}

\author{G.\ F. Lewis\altaffilmark{4}}

\author{Q.\ A. Parker\altaffilmark{1, 2, 5}}

\author{D.\ B. Zucker\altaffilmark{1, 2, 5}}

\author{N.\ F. Martin\altaffilmark{3}}

\author{A.\ W. McConnachie\altaffilmark{6}}

\author{M.\ J. Irwin\altaffilmark{7}}

\author{N. Tanvir\altaffilmark{8}}

\author{M.\ A. Fardal\altaffilmark{9}}

\author{A.\ M.\ N. Ferguson\altaffilmark{10}}

\author{S.\ C. Chapman\altaffilmark{7}}

\author{D. Valls-Gabaud\altaffilmark{11}}

\altaffiltext{1}{Department of Physics \& Astronomy, Macquarie University, NSW 2109, Australia.}
 
\altaffiltext{2}{Research Centre in Astronomy, Astrophysics and Astrophotonics (MQAAAstro), Macquarie University, NSW 2109, Australia.}
 
\altaffiltext{3}{Observatoire Astronomique, Universite« de Strasbourg, CNRS, 67000 Strasbourg, France.}
 
\altaffiltext{4}{Sydney Institute for Astronomy, School of Physics, A28, University of Sydney, 
Sydney NSW 2006, Australia.} 
 
\altaffiltext{5}{Australian Astronomical Observatory, PO Box 296, Epping, NSW 2121, Australia.}
 
\altaffiltext{6}{NRC Herzberg Institute of Astrophysics, 5071 West Saanich Road, Victoria, British Columbia, Canada V9E 2E7.}

\altaffiltext{7}{Institute of Astronomy, University of Cambridge, Madingley Road, Cambridge CB3 0HA, UK.}

\altaffiltext{8}{Department of Physics and Astronomy, University of Leicester, Leicester LE1 7RH, UK.}

\altaffiltext{9}{University of Massachusetts, Department of Astronomy, LGRT 619-E, 710 N. Pleasant Street, Amherst, Massachusetts 01003-9305, USA.}

\altaffiltext{10}{Institute for Astronomy, University of Edinburgh, Royal Observatory, Blackford Hill, Edinburgh EH9 3HJ, UK.}

\altaffiltext{11}{Observatoire de Paris, LERMA, 61 Avenue de l'Observatoire FR 75014 Paris
France.}
 

\begin{abstract}
 In `A Bayesian Approach to Locating the Red Giant Branch Tip Magnitude (PART I),' a new technique was introduced for obtaining distances using the TRGB standard candle. Here we describe a useful complement to the technique with the potential to further reduce the uncertainty in our distance measurements by incorporating a matched-filter weighting scheme into the model likelihood calculations. In this scheme, stars are weighted according to their probability of being true object members. We then re-test our modified algorithm using random-realization artificial data to verify the validity of the generated posterior probability distributions (PPDs) and proceed to apply the algorithm to the satellite system of M31, culminating in a 3D view of the system. Further to the distributions thus obtained, we apply a satellite-specific prior on the satellite distances to weight the resulting distance posterior distributions, based on the halo density profile. Thus in a single publication, using a single method, a comprehensive coverage of the distances to the companion galaxies of M31 is presented, encompassing the dwarf spheroidals Andromedas I - III, V, IX-XXVII and XXX along with NGC147, NGC 185, M33 and M31 itself. Of these, the distances to Andromeda XXIV - XXVII and Andromeda XXX have never before been derived using the TRGB. Object distances are determined from high-resolution tip magnitude posterior distributions generated using the Markov Chain Monte Carlo (MCMC) technique and associated sampling of these distributions to take into account uncertainties in foreground extinction and the absolute magnitude of the TRGB as well as photometric errors. The distance PPDs obtained for each object both with, and without the aforementioned prior are made available to the reader in tabular form. The large object coverage takes advantage of the unprecedented size and photometric depth of the Pan-Andromeda Archaeological Survey (PAndAS). Finally, a preliminary investigation into the satellite density distribution within the halo is made using the obtained distance distributions. For simplicity, this investigation assumes a single power law for the density as a function of radius, with the slope of this power law examined for several subsets of the entire satellite sample. 
\end{abstract}



 
\section{Introduction}
\label{s_intro}

The Tip of the Red Giant Branch (TRGB) is a well established standard candle for ascertaining distances to extended, metal poor structures containing a sufficient red giant population. Its near constant luminosity across applicable stellar mass and metallicity ranges (see \citealt{Iben83}) arises due to the prevailing core conditions of these medium-mass stars as core helium fusion ensues. Their cores  lack the necessary pressure to ignite immediate helium fusion on the depletion of their hydrogen fuel and so they continue to fuse hydrogen in a shell around an inert, helium ash core. This core is supported by electron degeneracy, and grows in mass as more helium ash is deposited by the surrounding layer of hydrogen fusion. On reaching a critical mass, core helium fusion ignites, and the star undergoes the helium flash before fading from its position at the tip of the Red Giant Branch, to begin life as a Horizontal Branch star. Due to the very similar core properties of the stars at this point, their energy output is almost independent of their total mass, resulting in a distinct edge to the RGB in the Color-Magnitude Diagram (CMD) of any significant red giant population. 

With the TRGB standard candle applicable wherever there is an RGB population, it is an obvious choice for obtaining distances to the more sparsely populated objects in the Local Group and other nearby groups where Cepheid Variables seldom reside. Even when Cepheids are available, the TRGB often remains a more desirable alternative, requiring only one epoch of observation, and facilitating multiple distance measurements across an extended structure. Good agreement between TRGB obtained distances, and those obtained using Cepheid Variables as well as the much fainter RR Lyrae Variables have been confirmed by \citet{Salaris97}, with discrepancies of no more than $\sim 5 \%$ (see also \citealt{Tammann10} for an extensive list of distance comparisons utilizing the three standard candles). Of the satellites of M31, many are very faint and poorly populated and thus have poorly constrained distances which propagate on into related measurements concerning the structure of the halo system. Hence, a technique for refining the distances that can be applied universally to all halo objects, whilst \emph{accurately} conveying the associated distance errors, has been a long sought goal.  

In `A Bayesian Approach to Locating the Red Giant Branch Tip Magnitude (PART I)' - \citet{Conn11}, hereafter Paper I, we reviewed the challenges of identifying the TRGB given the contamination to the pure RGB luminosity function (LF) typically encountered. We also outlined some of the methods that have been devised to meet these challenges since the earliest approach, put forward by \citet{Lee93}. We then introduced our own unique Bayesian Approach, incorporating MCMC fitting of the LFs. This approach was essentially the base algorithm, designed to easily incorporate priors to suit the task at hand. Here we present the results of an adaptation of that algorithm, intended for use on small, compact objects - specifically the dwarf spheroidal companions of M31. Once again, we utilize the data of the Pan-Andromeda Archaeological Survey \citep{McConn09}, a two-color ($i' = 770$ nm, $g' = 487$ nm) panoramic survey of the entire region around M31 and M33 undertaken using the Canada-France-Hawaii Telescope (CFHT). The tip is measured in $i'$ band where dependance on metallicity is minimal. Following a recap of the base method in \S \ref{recap}, we introduce the aforementioned new adaptations to the method in \S \ref{ss_MF_description} and in \S \ref{ss_MF_tests} we describe the results of tests intended to characterize the modified algorithms performance as well as check the accuracy of its outputs. In addition, \S \ref{ss_add_prior} outlines the application of a further prior on the satellite distances. \S \ref{ss_galaxy_distances} presents the results of applying the modified algorithm to the companions of M31, while  \S \ref{M31_Dist} details the method by which the object-to-M31 distances are obtained and \S \ref{ss_halo_analysis} uses the obtained distances to analyze the density profile of these objects within the halo. Conclusions follow in \S \ref{s_Conclusions}.

 
 \section{A recap of the Base Method}
 \label{recap}
 
 In Paper I, we introduced our `base' method, whereby the LF of a target field was modeled by a single, truncated power law (the RGB of the object of interest) added to a representative background polynomial. The location of the truncation (the TRGB) and the slope of the power law were set as free parameters of the model, with the best fit derived using an MCMC algorithm. The functional form of the background component was modeled by directly fitting a polynomial to the LF of an appropriate background field, and then scaling the polynomial to reflect the expected number of background stars in the target field. The resulting model was then convolved with a Gaussian of width increasing in proportion to the photometric error as a function of magnitude. The posterior distribution in the tip magnitude returned by the MCMC, which thus already incorporates the photometric error, is then sampled together with Gaussian distributions representing the distribution in the absolute magnitude of the tip ($M^{TRGB}_{i} = -3.44 \pm 0.05$) and the distribution in the extinction ($A_{\lambda} \pm 0.1 A_{\lambda}$) to give a final posterior distribution in the distance. The mode of this distribution is then adopted as the distance to the object, with the $\pm 1 \sigma$ error calculated from the portion of the distribution lying on the far and near side of the mode respectively.   
 
A more detailed discussion of the assumptions and rationale behind the base method is provided in paper I , but the reader should again be made aware of the most fundamental assumptions it entails. At the heart of the calculations of course is the choice of the absolute magnitude of the tip and its associated uncertainty. We adopt the values of this parameter stated above based on the value derived for the SDSS i-band in \citet{Bellazzini08}, noting the near-identical bandpass characteristics of the MegaCam i-band filter as detailed by \citet{Gwyn10}. We adopt somewhat smaller uncertainties than those derived by \citet{Bellazzini08} following the same argument as \citet{McConn04} that the quoted uncertainty in the absolute magnitude of the tip is conservative, and it is a systematic error effecting all distance measurements in an identical way. As almost all applications of the distances to the satellites are concerned with their relative positions to one another and M31, this component of the error is of minimal importance. Nevertheless, it often forms the major component of the quoted errors in our distances. 
 
 Mention should also be made as to the effects of metallicity and internal reddening within the objects under study as well as the zero-point uncertainty in the PAndAS photometry. Whilst there is a metallicity dependence of  $M^{TRGB}_{i}$, (though minimal when compared with other bands), it is only really an issue for more metal-rich targets (e.g. $[Fe/H] > -1$, see \citealt{Bellazzini08} Fig. 6) and thus will primarily effect measurements to the large, diverse systems such as M31 itself and M33. But the TRGB for more metal rich populations is fainter than that for their metal-poor counterparts and thus it is this metal-poor population component which dominates the measurement. A similar situation is encountered with the internal reddening present in the objects under study, where the vast majority of objects, chiefly the dwarf spheroidal galaxies, are almost completely devoid of such effects. Those objects most strongly effected are the large, well populated systems which will provide ample signal from the least effected stars on the near side of the system, for a good distance determination.  The uncertainty in the zero-point of the photometry is consistent throughout the survey at approximately 0.02 magnitudes \citep{Ibata12}.
 
 Lastly, a brief discussion of the distance posterior distributions themselves is warranted. As noted above, they are produced by the sampling of the distribution of possible tip positions (as generated by the MCMC and with photometric errors incorporated) along with sampling of the Gaussian distributions representing the uncertainties in the foreground extinction ($A_{\lambda}$) and in the absolute magnitude of the tip ($M^{TRGB}_{i}$). Specifically, $500,000$ possible distances are drawn to form the distance PPD, where for each draw $\kappa$, the distance modulus $\mu$ is:
 
 \begin{equation}
 	\mu(\kappa) = m^{TRGB}_{i}(\kappa) - A_{\lambda}(\kappa) - M^{TRGB}_{i}(\kappa).
	\label{e_dist_samp}
\end{equation}
where each of $m^{TRGB}_{i}(\kappa)$, $A_{\lambda}(\kappa)$ and $M^{TRGB}_{i}(\kappa)$ are the values drawn from the uncertainty distributions in the tip position, foreground extinction and absolute magnitude of the tip respectively. The foreground extinction and its uncertainty varies from object to object but the error in the absolute magnitude of the tip is a systematic error as already discussed. In using this method, there are two situations that can be encountered. The first  is that the object is very well populated and the tip position is thus well constrained with a narrow PPD. In such instances, the uncertainty in $M^{TRGB}_{i}$ far outweighs any other contributions to the error budget and is almost solely responsible for the width of the distance PPD. In the second situation, the object is poorly populated and the tip position PPD is very wide and typically asymmetric. If the LF population is not extremely low, the uncertainty in $M^{TRGB}_{i}$ will contribute noticeably to the distance PPD, otherwise the distance PPD will essentially depend solely on the uncertainty in the determined tip positions. Hence whilst some of the smaller contributions to the distance uncertainties are omitted from the calculations, their over all effects will be washed out by the contributions from these two principle sources of error.

 
\section{Addition of a Matched Filter}
\label{s_MF}

\subsection{Matched Filtering using Radial Density Profiles}
\label{ss_MF_description} 

With the introduction of our method in Paper I, it was stressed that one of its greatest attributes was its adaptability to the prior knowledge available for the object of interest. When applying the method to compact satellites, there is one very conspicuous attribute that can be incorporated into the prior information constraining the model fit - namely, the object's density as a function of radius. The simplest way to achieve this is with the addition to the algorithm of a matched-filter weighting scheme, wherein the weighting is \emph{matched} to the specific data by accounting for the data within the filter itself.  

The successes of \citet{Rockosi02} using a matched filter in colour-magnitude space to identify member stars of globular cluster Palomar 5 amidst the stellar background provide the inspiration for our technique. They make use of the characteristic RGB of the globular cluster to weight stars as to their likelihood of being cluster members. To achieve such a goal, a matched filter can be created by binning the CMD of the field in which the cluster lies into a 2D matrix and then dividing that matrix by a similarly created background matrix. Stars found in the densest regions of the resulting matched filter CMD are then assigned the highest weight, being the most likely cluster members. In this way, they can greatly improve the signal-to-noise ratio (SNR) with respect to that of their original, unmodified data and are able to trace tidal streams from the globular cluster well into the surrounding background. Hence we have applied a similar approach to weight field stars fed to the MCMC in terms of their probability of being object members. In our case however, the stars proximity to the object's center provides the basis for the weighting scheme, with the inner most stars being the most likely to be actual object members as opposed to background stars, and so a one-dimensional matched filter is sufficient.

The first step in implementing our weighting scheme is to ascertain a model of stellar density as a function of radius specific to the object of interest. For this purpose, we employ the best fits presented in \citet{Martin12} for the dwarf spheroidal satellites, wherein the optimal ellipticity $\epsilon$, position angle (PA), half-light radii ($r_h$) and object centers are given for exponential density profiles fitted to each satellite. For the two dwarf ellipticals, in the case of NGC147 we assume  $\epsilon = 0.44$ and $PA = 28^{\circ}$ as  specified by \citet{Geha2010} and we derive the $r_h$ manually as $10'$, which produces the best fit profile to the data when coupled with the other 2 parameters. For NGC185, we adopt $\epsilon = 0.26$ and $PA = 41^{\circ}$ based on the findings of \citet{Hodge1963} and once again derive the $r_h$ manually, this time as $6'$. For both NGC147 and 185, we employ the object centers derived from the 2 Micron All Sky Survey (2MASS - \citealt{2MASS}). With the ellipticity, position angle, half-light radius and object center know, we can proceed to produce a weighting scheme proportional to the density profile $\rho$ of the object, where $\rho$ is of the form:

\begin{equation}
	\rho(r_{\epsilon}) = e^{\frac{-r_{\epsilon}}{R}} 
	\label{e_weight1}
\end{equation}
where $R = \frac{r_h}{1.678}$ is the scale radius and $r_{\epsilon}$ is the elliptical radius at which the star lies, as now defined. With the PA and object center of the object known, a rotation of coordinates is used to define each star's position $(x',y')$ with respect to the center of the ellipse. The projected elliptical radius $r_{\epsilon}$ of the ellipse on which the star lies is then:

\begin{equation}
	r_{\epsilon} = \sqrt{(y')^2 + \left( \frac{x'}{1-\epsilon} \right)^2}
	\label{e_weight2}
\end{equation} 
where the $y'$ axis is assumed as the major axis of the ellipse. 

Whilst Eq. \ref{e_weight1} gives us the functional form of our weighting scheme, it is further necessary to define the absolute values of the weights given to each star, so as to scale them appropriately with respect to the background density $\rho_{bg}$. This is achieved by insuring that the area under the function $\rho(r_{\epsilon})$ between any imposed inner and outer radius limits is set equal to the number of signal stars in the observed region. Hence, our weighting scheme is ultimately defined by:

\begin{equation}
	W(r_{\epsilon}) = Se^{\frac{-r_{\epsilon}}{R}} 
	\label{e_weight3}
\end{equation}
with,

\begin{equation}
	S =  \frac{(\rho_{total} - \rho_{bg}) \times A}{2 \pi R(1-\epsilon) [(e^{\frac{-r_{inner}}{R}})(R + r_{inner}) - (e^{\frac{r_{outer}}{R}})(R + r_{outer})]}
	\label{e_weight4}
\end{equation} 
where $\rho_{total}$ is the density of stars in the observed region before subtraction of the background density, $A$ is the area of the observed region which is either an ellipse in the (usual) case that $r_{inner} = 0$ or an elliptical annulus otherwise. $r_{inner}$ and $r_{outer}$ are the inner and outer cutoffs respectively of the range of $r_{\epsilon}$ values observed. 

\begin{figure}[htbp]
\begin{center}
\includegraphics[width = 0.35\textwidth,angle=-90]{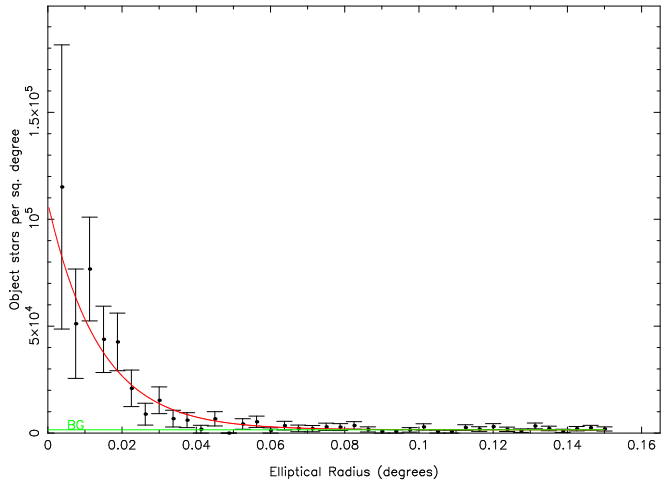}				  
\caption{Radial Density Profile (proportional to Object Membership Probability) for Andromeda X. The error bars represent the Poisson Error in the density for each bin, with each bin representing an elliptical annulus at the stated radius. Hence the inner most annuli have the smallest areas and thus the largest error bars. Note that this binned density distribution is for comparison only and has no bearing on the fit. The background level is marked `BG.'}
\label{ANDX_RDP}
\end{center}

\end{figure}

In Fig. \ref{ANDX_RDP}, the result of our fitting procedure as applied to the sparsely populated dwarf spheroidal Andromeda X is presented. In this case, stars out to $r_{\epsilon} = 0.15^{\circ}$ are fitted, with no inner cutoff radius imposed. Whilst most of the satellites are too poorly populated for blending to be an issue, in the case of several, the stellar density counts at the inner most radii drop off in spite of the predicted counts from the fitted density profile. This is a good indicator of blending or overcrowding in those radii which can hinder the accuracy of the photometry for the affected stars and so in such cases, these inner radii are omitted. This was the case with Andromeda III ($r_{inner} =  0.0175^{\circ}$), Andromeda V ($r_{inner} =  0.011^{\circ}$) and Andromeda XVI ($r_{inner} =  0.005^{\circ}$). For the dwarf ellipticals NGC147 and NGC185, it was found beneficial to avoid the inner regions all together, with the presence of a wider range of metallicities in these regions degrading the contrast of the RGB tip. Similarly, an outer cutoff radius was chosen for these objects inside of 3 $r_h$ to help sharpen the tip discontinuity, so that for NGC147, $r_{inner} =  0.28^{\circ}$ and $r_{outer} =  0.33^{\circ}$ and for NGC185, $r_{inner} =  0.18^{\circ}$ and $r_{outer} =  0.26^{\circ}$.  M31 and M33 are treated similarly to the dwarf ellipticals but with still thinner annuli so that any weighting is unnecessary. They are discussed in more detail in \S \ref{ss_galaxy_distances}.

With regard to the actual likelihood calculations used at each iteration of the MCMC, these are undertaken not by simple multiplication of the likelihood for each star by the respective weight, but by physically adjusting the relative proportions of the RGB and background components of the luminosity function. Up until now, we have assumed a generic LF and calculated the likelihood contributions from each star from this single LF. But in reality, the outer regions of the field are more accurately represented by a shallow-signal/ high-background LF while the innermost stars obey a LF which has almost no background component. Hence using the radial density profile obtained above, we can essentially build an individual LF for each star, tailored to suit its position within the object. In practice, this is achieved with almost no extra computational effort, as the background and signal can be normalized separately and only the signal component is changed by the MCMC at each iteration so that the background component need only be generated once. The two components are normalized to contain an area of unity and then the bin of each corresponding to the star's magnitude is scaled according to the ratios of the star's weight and the background level when its contribution to the model likelihood is calculated by the MCMC.   

\begin{figure}[htbp]
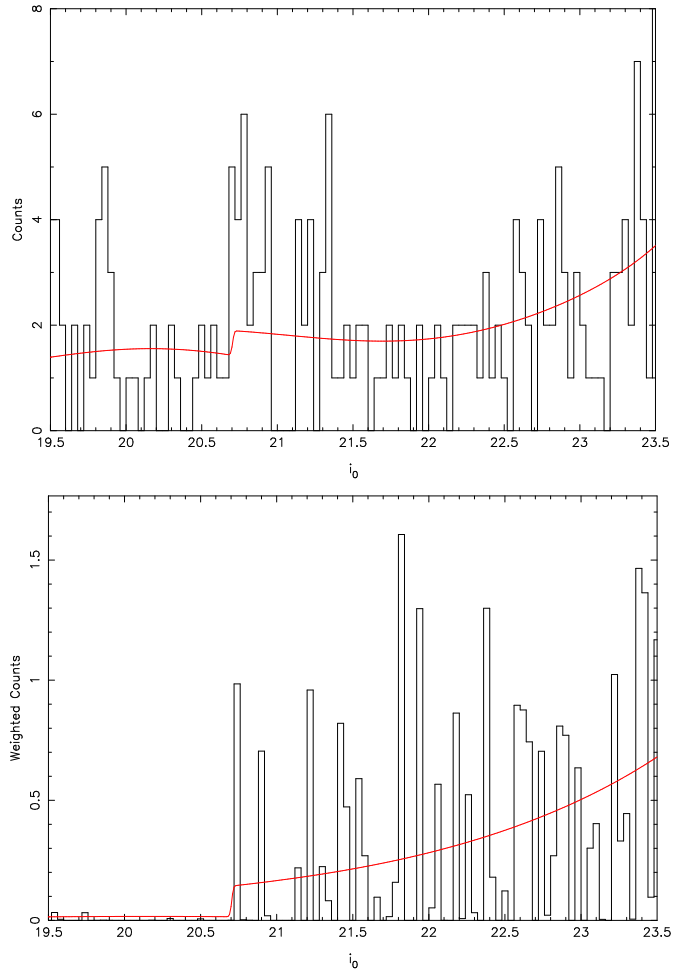

\begin{center} 
$ \begin{array}{c} 
\includegraphics[width = 0.35\textwidth,angle=-90]{Figures/ANDX_LF.ps} 
\\ \includegraphics[width = 0.35\textwidth,angle=-90]{Figures/ANDX_wLF.ps}	
\end{array}$
\end{center}
\caption{Best fit model to the luminosity function of Andromeda X, obtained with the addition of matched filtering. The top figure shows the best fit overlaid on the un-modified LF (i.e. histogram created without the weighting afforded by the matched filter). The bottom figure shows the same best fit model after applying the weighting. A field radius of $0.15^{\circ}$ was used to generate the LF histograms, wherein each star contributes between 0 and 1 `counts', depending on its proximity to the field centre and the density profile of the object.}
\label{ANDX_LF_fits}

\end{figure}

\begin{figure}[htbp]
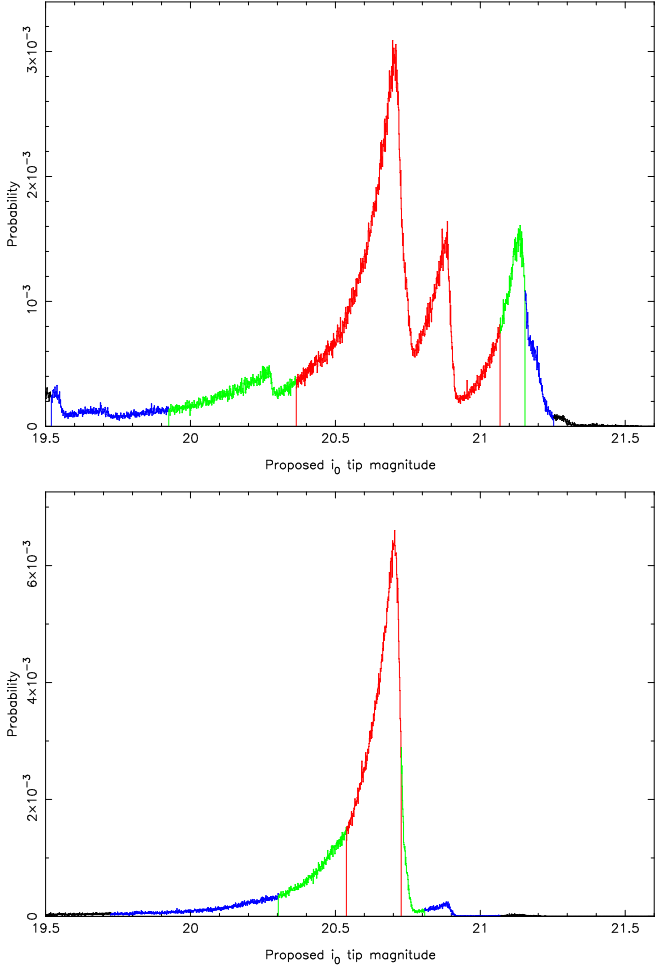

\begin{center} 
$ \begin{array}{c} 
\includegraphics[width = 0.35\textwidth,angle=-90]{Figures/ANDX_PPD_nmf.ps} 
\\ \includegraphics[width = 0.35\textwidth,angle=-90]{Figures/ANDX_mag_PPD.ps}	
\end{array}$
\end{center}
\caption{Posterior distributions obtained for Andromeda X before (top) and after (bottom) the application of the matched filter. For the `before' case, a circular field of radius $0.05^{\circ}$ $(2.143 \times r_h)$ has been chosen, specifically to provide the most possible signal with the least possible background contamination. For the `after' case, the same LF as presented in Fig. \ref{ANDX_LF_fits} is used.}
\label{ANDX_PPDs}

\end{figure}

The result of the incorporation of this extra prior information is a marked improvement in the performance of the algorithm for the more sparsely populated targets. In such objects, the RGB component is typically overwhelmed by non-system stars, even with the most carefully chosen field size. This can greatly diminish the prospects of obtaining a well constrained tip measurement. This is apparent from figures \ref{ANDX_LF_fits} and \ref{ANDX_PPDs} which show the luminosity function and corresponding posterior distributions before and after the application of the matched filter to the dwarf spheroidal Andromeda X. With the matched filtering applied, the great majority of non-system stars are severely suppressed, revealing clearly the RGB component, which in turn provides much stronger constraints on the location of the tip, as evidenced by Fig. \ref{ANDX_PPDs}. Herein lies an example of the power of the Bayesian approach, where a single prior can cast the available data in a completely different light.


\subsection{A Test for the refined algorithm}
\label{ss_MF_tests}

In \S2.3 of Paper I, the results of a series of tests were presented that characterized the performance of our original algorithm given a range of possible background density levels and LF populations. Here we present the results of similar tests applied to our new, matched-filter equipped algorithm, but with some important differences. Most fundamentally, the way our artificial test data is generated is quite different. As we are now concerned with the position of each star in the field, a distance from field centre must be generated for each star. To do this, we have randomly assigned a radial distance to each star, but weighted by a circularly symmetric ($\epsilon = 0$) exponential density profile . Further to this, the magnitudes of our stars are now generated directly from our convolved LF, so that photometric error as a function of stellar magnitude is incorporated.

\begin{figure}[htbp]
\begin{center}
\includegraphics[width = 0.35\textwidth,angle=-90]{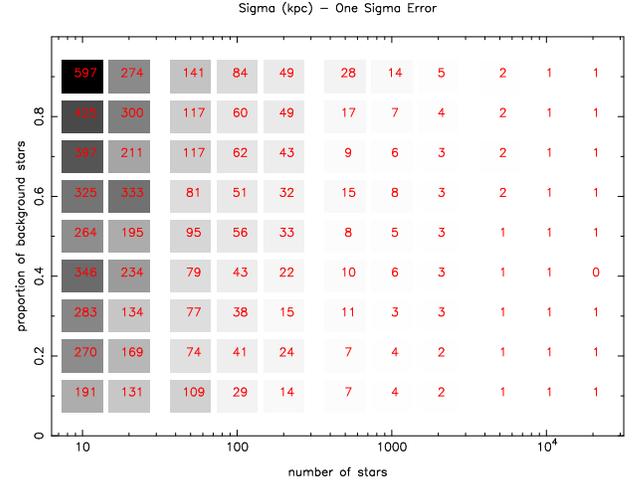}				  
\caption{A grey-scale map of the one-sigma error in tip magnitude obtained 
for different combinations of background height and number of sources. The 
actual value recorded for the error (in kpc) is overlaid on each pixel in red. Each value is the average of twenty 50000 iteration runs for the given background height/ LF population combination.}
\label{sigma}
\end{center}

\end{figure}

\begin{figure}[htbp]
\begin{center}
\includegraphics[width = 0.35\textwidth,angle=-90]{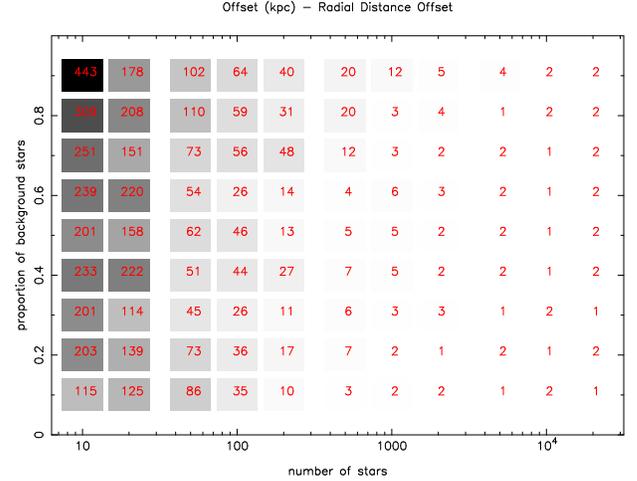}				  
\caption{A grey-scale map of the offset of the measured tip value from the true tip value obtained for different combinations of background height and number of sources. The actual value recorded (in kpc) is overlaid on each pixel in red. Each value is the average of twenty 50000 iteration runs for the given background height/ LF population combination.}
\label{offset}
\end{center}

\end{figure}

The other important change from the previous tests concerns the way in which the artificial luminosity functions are populated. Whereas in the former tests all of the sampled stars were drawn from the model LF within the one magnitude range $20 \le m_{star} \le 21$, in the current tests the stars are drawn from within the much larger magnitude range actually utilized for our satellite measurements, namely $19.5 \le m_{star} \le 23.5$. Hence a 100 star LF in these tests for example corresponds to a much smaller sample of stars than in the tests described in  \S2.3 of  Paper I. Aside from these critical differences, the current tests are undertaken and presented as per the previous publication, with measurements of the average sigma and tip offset  given for each combination of background level ($f$) .vs. number of stars ($ndata$) where $f = 0.1, 0.2, ..., 0.9$ and $ndata = 10, 20, 50, 100, 200, 500, 1000, 2000, 5000, 10000, 20000$. The results are presented in figures \ref{sigma} and \ref{offset} respectively.


Examination of the figures reveals the expected trend of increased one-sigma error and tip offset with increasing background height and decreasing LF population levels. Once again, there is very good agreement between the derived errors and the actual offsets obtained. Most importantly, it is clear by comparing these results with those of Paper I, that the Matched Filtering has greatly diminished the effects of the background contamination, as exemplified by the much gentler increase in $1\sigma$ errors and offsets with increasing background star proportion.  



\subsection{An additional Prior}
\label{ss_add_prior}

In addition to our density matched filter, a further prior may be devised so as to constrain our distance Posterior Probability Distributions (PPDs) in accordance with our knowledge of the M31 halo dwarf density profile. The expected fall off in density of subhaloes within an M31 sized galaxy halo is not well constrained.  The largest particle simulation of an M31 sized dark matter halo to date, the Aquarius Project \citep{Springel08}, favored the density of subhaloes to fall off following an Einasto Profile with $r_{-2} = 200$ kpc and $\alpha = 0.678$, and furthermore identified no significant dependence of the relationship on subhalo mass. For the specific case of the satellites within the M31 halo, \citet{Richardson11} found a relation of  $\rho \propto r^{-\alpha}$ where $\alpha = 1$ a better fit to the data, drawing largely from the PAndAS survey, although this does not take into account the slightly irregular distribution of the survey area. We adopt this more gentle density fall off with radius giving us a more subtle prior on the satellite density distribution and note that $\alpha$ may be changed significantly without great effect on our measured distances.

So in effect, we assume a spherical halo centered on M31, such that $\rho(sat) \propto r^{-1}$ and integrate along a path through the halo at an angle corresponding to the angular displacement on the sky of the satellite from M31. This yields an equation of the form:

\begin{equation}
	P(d) \propto \frac{d^2}{\sqrt{(d^2 + 779^2 - 2d \times 779 \times cos(\theta))^{\alpha}}}
	\label{e_dist_prior}
\end{equation}
where $\alpha = 1$, $779$ $kpc$ is the distance to M31 and $P(d)$ is the relative probability of the satellite lying at distance $d$ (in $kpc$) given an angular separation of $\theta$ degrees from M31. Note that this produces a peak where the line of sight most closely approaches M31, and that $P(d >> 779)$ is approximately proportional to $d$. The equation is normalized between limits appropriate to the size of the halo.

\begin{figure}[htbp]
\begin{center}
\includegraphics[width = 0.35\textwidth,angle=-90]{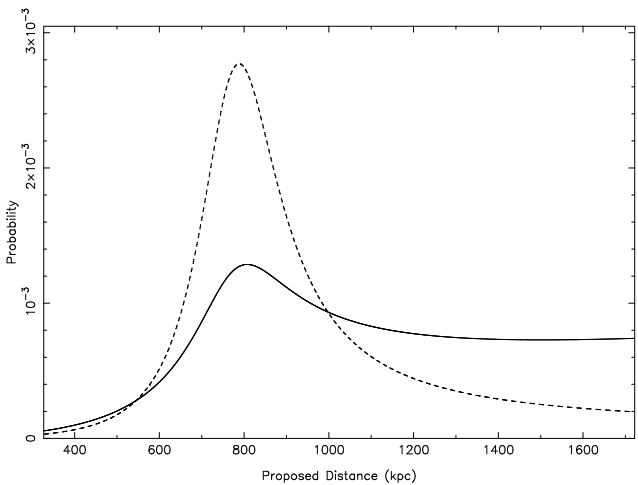}				  
\caption{The Distance Prior applied to Andromeda XIII (solid line; $\alpha = 1$). The distribution gives the likelihood of the satellite existing at a particular distance, given an angular separation on the sky of $8.5^{\circ}$ from M31 (the halo centre), and assuming a distance of 779 kpc for M31. The distribution peaks where the line of sight traverses the inner most region of the halo, and flattens out at large distances due to the increasing volume of the halo subtended by the unit of solid angle observed. The same prior with $\alpha = 2$ is shown as a dashed line for reference. Whilst this value for alpha is in closer agreement with the results of \S \ref{ss_halo_analysis}, we deliberately adopt the less restrictive $\alpha = 1$ prior, so as not to suppress the probability of satellites in the outer halo too greatly.} 
\label{DistPrior}
\end{center}

\end{figure}

We thus generate a separate prior for the probability as a function of distance for each satellite, tailored to its specific position with respect to M31. The effect of the prior is to suppress unlikely peaks in the multi-peaked posterior distributions obtained for certain satellites, while leaving the peak positions unaffected. As such, the prior has very little effect on single peaked distributions, whatever the angular position and distance of the satellite it represents. The distance prior applied to the Andromeda XIII distance PPD is shown in Figure \ref{DistPrior} for illustration.


\section{A New Perspective on the Companions of M31}
\label{s_results}

\subsection{Galaxy Distances}
\label{ss_galaxy_distances}

The PAndAS Survey provides us with a unique opportunity to apply a single method to a homogeneous data sample encompassing the entire M31 halo out to 150 kpc. The data encompasses many dwarf spheroidals, along with the dwarf ellipticals NGC147 and NGC185, and of course the M31 disk itself with additional fields bridging the gap out to the companion spiral galaxy M33, some $15^{\circ}$ distant. Of these objects, the vast majority have metallicities $[Fe/H] \leq -1$, so that any variation in the absolute magnitude of the tip is slight. Indeed, \citet{Bellazzini08} suggests that for such metallicities, the variation in the region of the spectrum admitted by the CFHT $i'$ filter is perhaps less than in Cousins' I. Perhaps of greatest concern are the cases of M31 and M33, which will contain substructure at a variety of metallicities. In this case however, the more metal rich portions will exhibit a fainter TRGB than those in the regime $[Fe/H] \leq -1$, such that the brightest RGB stars will fall within this regime. 

In this section we present distance measurements to these many halo objects, culminating in Fig. \ref{M31in3D}, a three-dimensional map of the satellite distribution, and Table \ref{sat_table}, which presents the satellite data pertinent to our distance measurements. Figures \ref{Dist_PPDs1} and \ref{Dist_PPDs2} present the distance posterior distributions obtained for every object in this study. It has been common practice in the majority of TRGB measurements to quote simply the most likely distance and estimated $1\sigma$ uncertainties, but this throws away much of the information, except in the rare case that the distance distribution is actually a perfect Gaussian. On account of this, as well as providing the actual distance PPDs themselves for visual reference, we also provide the same information in condensed tabular form, where the object distance is given at $1 \%$ increments of the PPD, both for the prior-inclusive cases (as in Figures \ref{Dist_PPDs1} and \ref{Dist_PPDs2}) and for the case in which no prior is invoked on the halo density. Note that for M31, no halo density prior is applied and so this column is set to zero. A sample of this information, as provided for Andromeda I, is presented in Table \ref{ppd_table}. The reader may then sample from these distributions directly rather than use the single quoted best fit value, thus taking into account the true uncertainties in the measurements. 

\begin{table}[ht] 

\caption{Tabulated Distance Posterior Distribution:}

\vspace{2.5 mm}

\centering      
\begin{tabular}{c c c}  
\hline\hline                        
Percentage & Distance (kpc - no density prior) & Distance (kpc) \\ [0.5ex] 
\hline                    
$1$ & $684$ & $687$ \\  [1.0ex]
$2$ & $688$ & $692$ \\  [1.0ex]
$3$ & $691$ & $695$ \\  [1.0ex]
$4$ & $693$ & $697$ \\  [1.0ex]
$5$ & $695$ & $699$ \\  [1.0ex]
$6$ & $697$ & $701$ \\  [1.0ex]
$7$ & $698$ & $702$ \\  [1.0ex]
$8$ & $699$ & $703$ \\  [1.0ex]
$9$ & $700$ & $704$ \\  [1.0ex]
$10$ & $701$ & $705$ \\  [1.0ex]
$:$ & $:$ & $:$ \\ [1.0ex]
$:$ & $:$ & $:$ \\ [1.0ex]
$100$ & $820$ & $820$ \\  [1.0ex]
\hline                    

\end{tabular} 

\label{ppd_table}  

\begin{flushleft} Distance posterior probability distributions for Andromeda I given at $1\%$ intervals for the case of no halo density prior (column 2) and with the angle-specific prior outlined in \S \ref{ss_add_prior} applied (column 3). \end{flushleft}

\end{table} 

Due to the large number of objects studied, it is not practical to discuss each in detail within this publication. For this reason, Andromeda I will be discussed in further detail below as a representative example, followed by two of the more problematic cases for completeness. Firstly however, we describe the exceptional cases of M31 itself and M33.

M31 and M33, due to their large extent on the sky and the variety of substructure in their disks require a slightly different approach to that used for the other objects in this study. As was the case for NGC147 and NGC185, it was necessary to define a thin elliptical annulus so as to limit as much as possible the amount of substructure from other radii contaminating the LF. For both M31 and M33 such a thin annulus was used that any weighting with respect to the elliptical radius of the stars was trivial and so no weighting was used. For M31, an ellipticity of $0.68$ was adopted, with $PA = 37^{\circ}$. The inner and outer elliptical cutoff radii were set to $2.45^{\circ}$ and $2.5^{\circ}$ respectively. To check for any inconsistencies in the TRGB location across the whole annulus, it was divided up into NE, NW, SE and SW quarters and then the distance measured from each quarter, giving distances of $782^{+19}_{-19}$, $782^{+18}_{-18}$, $775^{+20}_{-18}$ and $781^{+19}_{-19}$ kpc respectively. It is tempting to associate the slightly lower distance to the SE quadrant with the effects on the LF of the Giant Stellar Stream, though the distance is still within close agreement with the other 3 quadrants, such that all 4 are perfectly consistent. Hence, the distance was remeasured using the whole annulus to give $779^{+19}_{-18}$ kpc. This is in good agreement both with the findings of \citet{McConn05} ($785^{+25}_{-25}$) utilizing the TRGB, and the more recent determination by \citet{Riess12} using Cepheid Variables ($765^{+28}_{-28}$). 

For M33, we employ an ellipticity of $0.4$ as used by \citet{McConn05}, but find a position angle of $PA = 17^{\circ}$ in closest agreement with the data. Inner and outer elliptical radii of $r_{inner} = 0.75^{\circ}$ and $r_{outer} = 0.9^{\circ}$ were adopted to give a very sharp discontinuity at the location of the tip. After applying an appropriate color-cut, the qualifying stars were fed into our algorithm to give a distance of $820^{+20}_{-19}$ kpc. This distance is in good agreement with that of $809^{+24}_{-24}$ kpc obtained by \citet{McConn05} and yields an M33 to M31 distance of $214^{+6}_{-5}$ kpc. It is interesting to note that a variety of quite different M33 distances exist in the literature, with derived distance moduli ranging from 24.32 (\citet{Brunt05} - 730 kpc - water masers) through 24.92 (\citet{Bonanos06} - 964 kpc - Detached Eclipsing Binaries). Indeed, the variety of standard candles utilized would suggest that M33 provides an ideal environment for calibrating the relative offsets between them. \citet{McConnThesis} suggests that the dispersion of M33 distances in the literature is tied to an inadequate understanding of the extinction in the region of M33. Most measurements, including those presented here, use the Galactic extinction values derived by \citet{Schlegel98}, although these do not account for extinction within M33 itself and are calculated via an interpolation of the extinction values for the surrounding region. Nevertheless, the elliptical annulus employed in our approach will act to smooth out the field-to-field variation that might exist between smaller regional fields.

\subsubsection{Andromeda I - Example of an ideal luminosity function}
\label{ANDI}
\begin{figure*}[htbp]
\begin{center}
		
\begin{overpic}[width = 0.65\textwidth,angle=-90]{Figures/ANDI_CMD.ps}	
\put(10,65){\huge a)}
\put(71,70){\includegraphics[width = 0.25\textwidth,angle=-90]{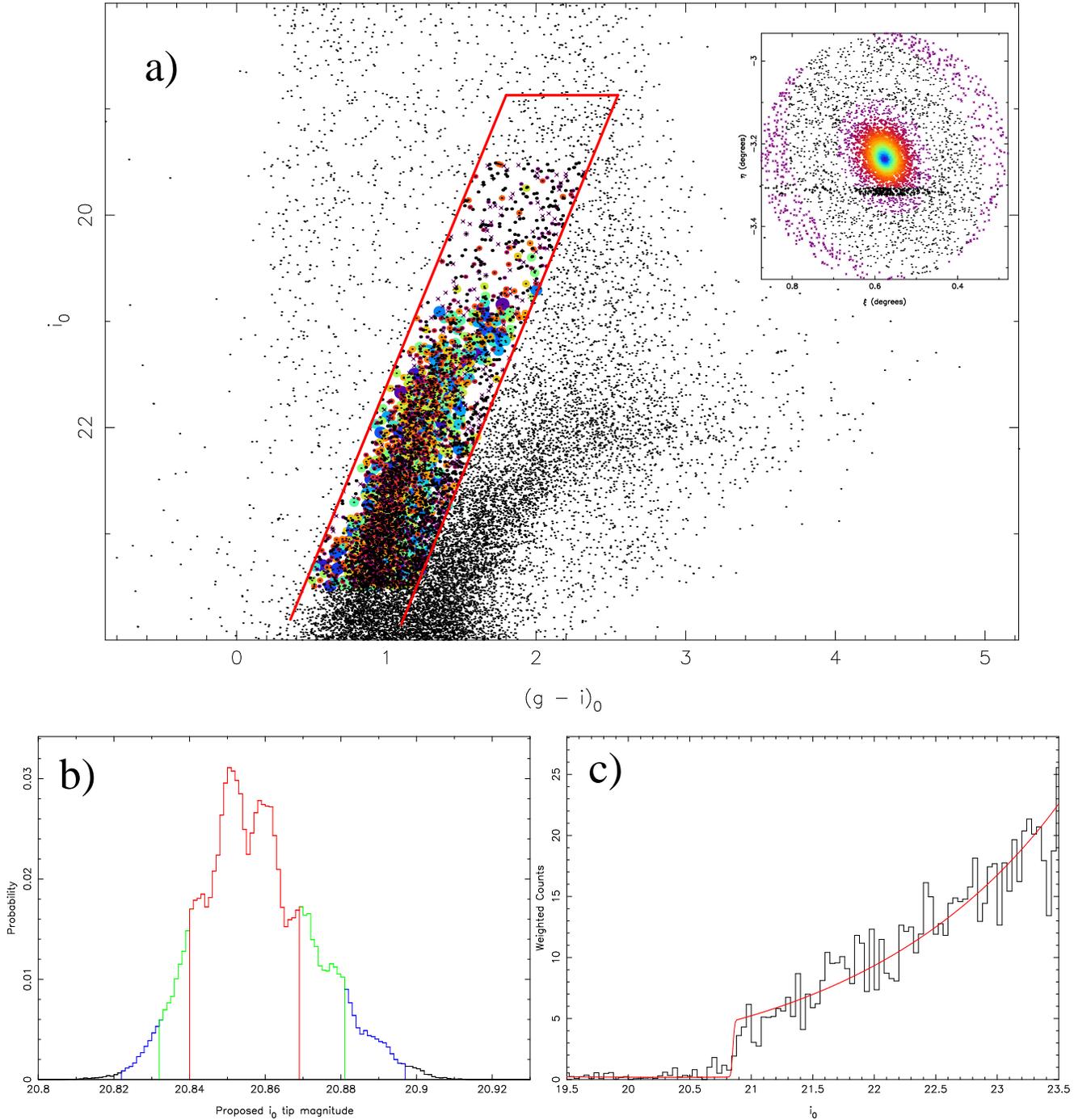}} 
\end{overpic}

$ \begin{array}{cc} 
\\ \begin{overpic}[width = 0.35\textwidth,angle=-90]{Figures/ANDI_mag_PPD.ps}
\put(10,63){\huge b)}
\end{overpic}
\ \begin{overpic}[width = 0.35\textwidth,angle=-90]{Figures/ANDI_LF.ps}
\put(10,63){\huge c)}
\end{overpic}

\end{array}$

\caption{Andromeda I: a): Color-coded CMD representing the weight given to each star in the field. Only stars within the red selection box with magnitudes $19.5 \leq i_0 \leq 23.5$ were fitted and hence color-coded. The second, fainter RGB lying toward the redder end of the CMD is that of the Giant Stellar Stream which passes behind our Andromeda I field. The inset at top right shows the field with the same color-coding and acts as a key. The field is divided into 20 radii bins following a linear decrease in density from the core (blue) to the field edge (purple). Stars marked as a purple `$\times$' lie outside of the outer elliptical cutoff radius $r_{outer}$. Stars marked as a black `$+$' are artificial stars used in the estimation of the background density and are ignored by the MCMC; b): Posterior Probability Distribution for the TRGB magnitude. The distribution is colour coded, with red indicating tip magnitudes within 68.2 \% (Gaussian 1-sigma) on either side of the distribution mode, green those within 90 \% and blue those within 99 \%; c): Weighted luminosity function of satellite with superimposed best-fit model in red. A star at the very centre of the satellite contributes 1 count to the luminosity function while those further out are assigned some fraction of 1 count in proportion with the satellite's density profile}

\label{fig_ANDI}

\end{center}

\end{figure*}

It would seem prudent to illustrate the performance of our new method by presenting the results for a range of the dwarf spheroidals from the most populated to the least populated. Hence Andromeda I, the first discovered and one of the two most highly populated of these objects is the obvious place to start. The field employed for our Andromeda I distance measurement incorporated stars at elliptical radii between $0^{\circ} \leq r_{\epsilon} \leq 0.3^{\circ}$ and, after removal of stars outside of the range $19.5 \leq i_0 \leq 23.5$ and beyond our chosen color-cut, yielded a star count of 4375. The CMD for this field is presented in Fig. \ref{fig_ANDI} a. This figure color-codes the stars in the CMD as per the color distribution in the inset field and plots them so that those innermost within the field (and hence those accorded the highest weight) are represented by the largest dots. In the case of Andromeda I, the RGB is so dominant over the background that our density matched filter is hardly necessary and hence does little to improve the already stark contrast. It is not surprising therefore that the distance and uncertainty obtained are almost identical to those obtained by the base method as presented in Paper I. Andromeda I is thus confirmed at a distance of $727^{+18}_{-17}$, which allows us to derive a similarly accurate separation distance from M31 of $68^{+22}_{-16}$ kpc. 

 
 \subsubsection{Andromeda XV - Example of a multi-peaked distance PPD}
\label{ANDXV}
\begin{figure*}[htbp]
\begin{center}
		
\begin{overpic}[width = 0.65\textwidth,angle=-90]{Figures/ANDXV_CMD.ps}	
\put(10,65){\huge a)}
\put(71,70){\includegraphics[width = 0.25\textwidth,angle=-90]{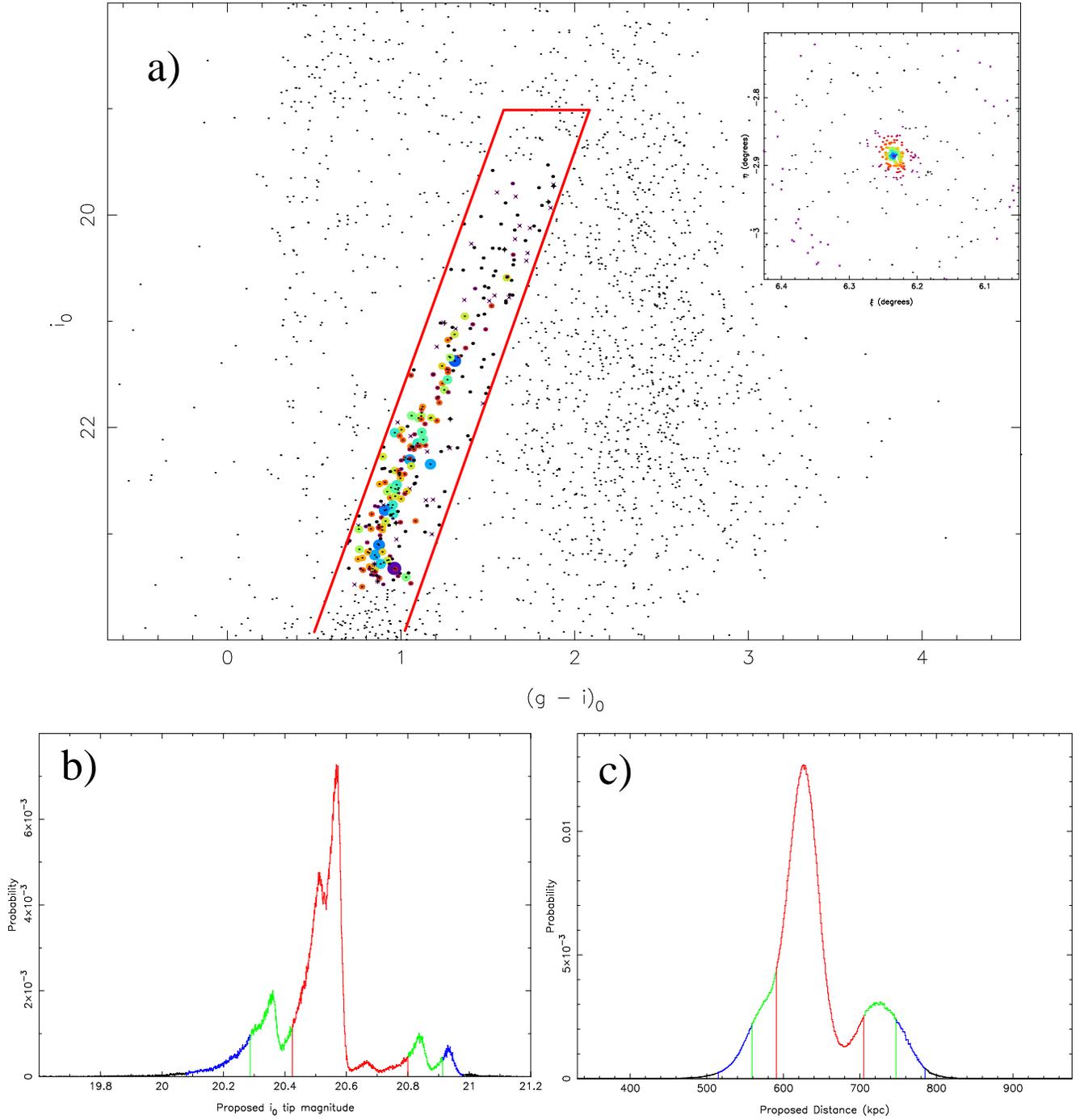}} 
\end{overpic}

$ \begin{array}{cc} 
\\ \begin{overpic}[width = 0.35\textwidth,angle=-90]{Figures/ANDXV_MagPPD.ps}
\put(10,63){\huge b)}
\end{overpic}
\ \begin{overpic}[width = 0.35\textwidth,angle=-90]{Figures/ANDXV_DistPPD.ps}
\put(10,63){\huge c)}
\end{overpic}

\end{array}$

\caption{Andromeda XV: a): Same as Fig. \ref{fig_ANDI} (a) but for Andromeda XV.; b): Same as Fig. \ref{fig_ANDI} (b) but for Andromeda XV; c): Sampled Distance Posterior Probability Distribution, obtained by calculating the distance 3 million times, each time randomly drawing on the tip magnitude, absolute magnitude of the tip and extinction from their respective probability distributions. The distribution is colour coded, with red indicating possible distances within 68.2 \% (Gaussian 1-sigma) on either side of the distribution mode, green those within 90 \% and blue those within 99 \%. Note that the large uncertainty in the absolute magnitude of the RGB tip is primarily responsible for the much smoother appearance of the distance PPD (c) compared with the tip PPD (b).}

\label{fig_ANDXV}

\end{center}

\end{figure*}

As an example of a dwarf spheroidal of intermediate size, we present the comparatively compact Andromeda XV. Far from being the tidiest example of the many intermediate sized objects covered in this study, Andromeda XV provides something of a challenge. Examination of Fig. \ref{fig_ANDXV} reveals a gradual rise in star counts when scanning from the top of the CMD color-cut faintward toward the Andromeda XV RGB and a correspondingly broad range in the possible tip locations in the tip magnitude PPD. Indeed, two peaks are prominent in the distance PPD of Fig. \ref{fig_ANDXV} c, with the distribution mode at 626 kpc (our adopted distance) and the $1\sigma$ credibility interval spanning from 591 kpc to 705 kpc as a consequence of the second peak. \citet{Ibata07} determine this object to lie at a distance of $630^{+60}_{-60}$ kpc, which would correspond to a tip magnitude of approximately $m^{TRGB}_{i} = 20.56$ assuming $M^{TRGB}_{i} = -3.44$. This is in excellent agreement with the $m^{TRGB}_{i} = 20.57^{+0.23}_{-0.14}$ recovered by this study. \citet{Letarte09} however derive a distance of $770^{+70}_{-70}$ kpc which places it toward the far edge of our $99\%$ credibility interval on the distance (see Fig. \ref{fig_ANDXV} c). This measurement was derived after 3 stars that had been found to lie close to the Andromeda XV RGB tip in the former investigation were identified as Galactic foreground stars, following measurements of their radial velocities obtained with the Deep Imaging Multi-Object Spectrograph on Keck II. Of these stars however, none lie within $2'$ from our object center, by which point the maximum possible weighting has already dropped to below $10\%$, meaning that even the highest weighted of these 3 stars will have minimal effect on the likelihood calculation. This would then suggest that each of these 3 stars have magnitudes consistent with belonging to the Andromeda XV RGB.


 \subsubsection{Andromeda XIII - Example of a very poorly populated luminosity function}
\label{ANDXIII} 
\begin{figure*}[htbp]
\begin{center}
		
\begin{overpic}[width = 0.65\textwidth,angle=-90]{Figures/ANDXIII_CMD.ps}	
\put(10,65){\huge a)}
\put(71,70){\includegraphics[width = 0.25\textwidth,angle=-90]{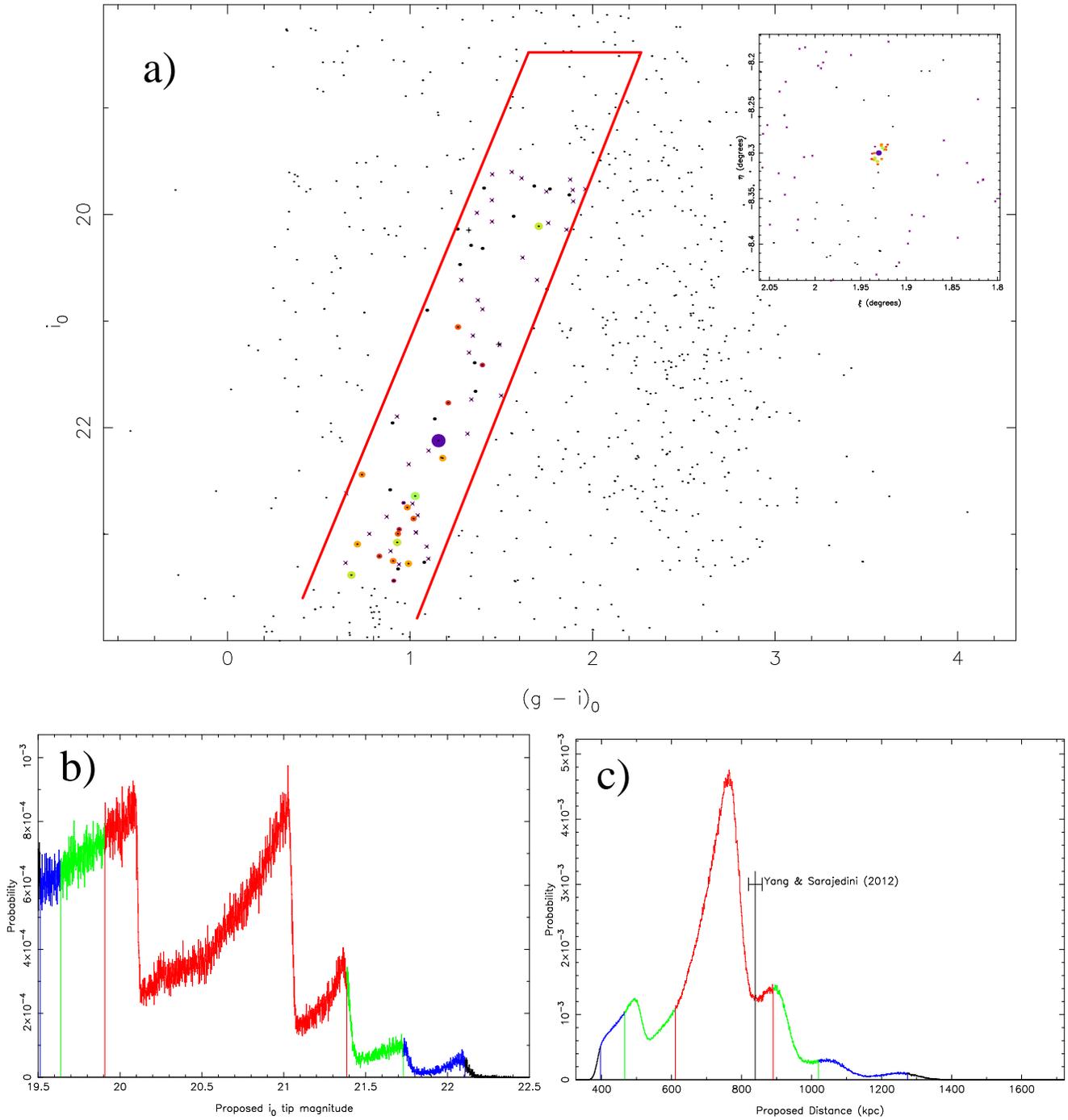}} 
\end{overpic}

$ \begin{array}{cc} 
\\ \begin{overpic}[width = 0.35\textwidth,angle=-90]{Figures/ANDXIII_MagPPD.ps}
\put(10,63){\huge b)}
\end{overpic}
\ \begin{overpic}[width = 0.35\textwidth,angle=-90]{Figures/ANDXIII_DistPPD.ps}
\put(10,63){\huge c)}
\end{overpic}

\end{array}$

\caption{Andromeda XIII: All figures as per Fig. \ref{fig_ANDXV}, but for Andromeda XIII. The distance derived by \citet{Yang12} is plotted in (c) along with error bars for comparison.}

\label{fig_ANDXIII}

\end{center}
\vspace{30 mm}
\end{figure*}

Andromeda XIII is among the most sparsely populated objects targeted by the current study and it is important to realize that it is impossible to obtain distances to such objects with small uncertainties using the TRGB standard candle, unless of course one of the few member stars can be positively identified as being right on the brink of core Helium fusion. Nevertheless, though large uncertainties are inevitable, an accurate estimation of those uncertainties is still achievable, and this is the aspiration of the method here presented. Distances to Andromeda XI and XIII have been obtained with higher accuracy using RR Lyrae stars as a Standard Candle with photometry from the Hubble Space Telescope \citep{Yang12}. In the case of Andromeda XI, the tip magnitude identified by our method agrees well with the distance identified by that study, but in the case of Andromeda XIII, a brighter star in the central regions of the field causes some confusion. Indeed in such a sparsely populated field it is quite difficult to apply any effective density-based weighting scheme. Nevertheless, after sampling the tip magnitude PPD (Fig. \ref{fig_ANDXIII} (b)), together with those for the absolute magnitude of the tip and the extinction in this region of sky to obtain a sampled distance PPD, and multiplying that distribution with the angle-specific halo density prior as is standard for all our measurements, we are able to produce a distance PPD (Fig. \ref{fig_ANDXIII} (c)) in good agreement with the findings of \citet{Yang12}.  

 
 
\begin{table*}[ht] 

\caption{M31 Satellite Parameters:}{Distance and associated parameters of M31 and its companions. All distance measurements utilize the data from the Pan-Andromeda Archaeological Survey \citep{McConn09}, and have been obtained using the method presented in this paper. A value of $M^{TRGB}_{i} = -3.44 \pm 0.05$ is assumed for the absolute magnitude of the RGB tip in CFHT MegaCam $i$-band, based on the value identified for the SDSS $i$-band \citep{Bellazzini08} and justified for use here by the color equations applicable to the new MegaCam $i$-band filter \citep{Gwyn10}. Values for the extinction in MegaCam $i$-band have been adopted as $A_{\lambda} = 2.086 \times E(B-V)$ for the same reasons, with uncertainties taken as $\pm 10 \%$. Note that the uncertainties in the M31 distance are based on the sampled distributions while the quoted value is that derived directly from the earth-distance as per Eq. \ref{e_CSR}. The last column gives alternative distances from the literature. TRGB derived distances are quoted wherever possible.} 

\vspace{5 mm}

\centering      
\begin{tabular}{c c c c c c}  
\hline\hline                        
Source & Distance Modulus & E(B-V) & Distance (kpc) & M31 Distance (kpc) & Literature Distance Values (kpc) \\ [0.5ex] 
\hline                    
M31 & $24.46^{+0.05}_{-0.05}$ & $0.062$ & $779^{+19}_{-18}$ & $-$ & $785^{+25}_{-25}$  TRGB; \citet{McConn05} \\  [1.0ex]
 &   &   &   &   &  $784^{+17}_{-17}$ RC; \citet{Stanek98} \\ [1.0ex]
 &   &   &   &   &  $765^{+28}_{-28}$ Ceph; \citet{Riess12} \\ [1.0ex]
\hline
And I & $24.31^{+0.05}_{-0.05}$ & $0.054$ & $727^{+18}_{-17}$ & $68^{+21}_{-17}$ & $731^{+18}_{-17}$ TRGB; \citet{Conn11}  \\ 
 &   &   &   &   &  $735^{+23}_{-23}$ TRGB; \citet{McConn04} \\ [1.0ex]
\hline 
And II & $24.00^{+0.05}_{-0.05}$ & $0.062$ & $630^{+15}_{-15}$ & $195^{+20}_{-17}$ & $634^{+15}_{-14}$ TRGB; \citet{Conn11}  \\ 
 &   &   &   &   &  $645^{+19}_{-19}$ TRGB; \citet{McConn04} \\ [1.0ex]
\hline     
And III & $24.30^{+0.05}_{-0.07}$ & $0.057$  & $723^{+18}_{-24}$ & $86^{+25}_{-15}$ & $749^{+24}_{-24}$ TRGB; \citet{McConn05} \\  [1.0ex]
\hline
And V & $24.35^{+0.06}_{-0.07}$ & $0.125$  & $742^{+21}_{-22}$ & $113^{+9}_{-6}$ &  $774^{+28}_{-28}$ TRGB; \citet{McConn05} \\  [1.0ex]
\hline
And IX & $23.89^{+0.31}_{-0.08}$ & $0.076$  & $600^{+91}_{-23}$ & $182^{+38}_{-66}$ &  $765^{+24}_{-24}$ TRGB; \citet{McConn05} \\  [1.0ex]
\hline
And X & $24.13^{+0.08}_{-0.13}$ & $0.126$  & $670^{+24}_{-39}$ & $130^{+60}_{-17}$ &  $667-738$ TRGB; \citet{Zucker07} \\  [1.0ex]
\hline
And XI & $24.41^{+0.08}_{-0.32}$ & $0.080$  & $763^{+29}_{-106}$ & $102^{+149}_{-1}$ &  $740-955$ TRGB; \citet{Martin06} \\ 
&   &   &   &   &  $735^{+17}_{-17}$ RR Ly; \citet{Yang12} \\ [1.0ex]
\hline 
And XII & $24.84^{+0.09}_{-0.34}$ & $0.111$  & $928^{+40}_{-136}$ & $181^{+19}_{-87}$ &  $825^{+85}_{-159}$ TRGB; (MCMC without MF) \\ 
&   &   &   &   &  $740-955$ TRGB; \citet{Martin06} \\ [1.0ex]
\hline 
And XIII & $24.40^{+0.33}_{-0.49}$ & $0.082$  & $760^{+126}_{-154}$ & $115^{+207}_{-2}$ &  $890^{+360}_{-361}$ TRGB; (MCMC without MF) \\ 
&   &   &   &   &  $740-955$ TRGB; \citet{Martin06} \\ [1.0ex]
&   &   &   &   &  $839^{+20}_{-19}$ RR Ly; \citet{Yang12} \\ [1.0ex]
\hline 
And XIV & $24.50^{+0.06}_{-0.56}$ & $0.060$  & $793^{+23}_{-179}$ & $161^{+81}_{-3}$ &  $630-850$ TRGB; \citet{Majewski07} \\  [1.0ex]
\hline
And XV & $23.98^{+0.26}_{-0.12}$ & $0.046$  & $626^{+79}_{-35}$ & $174^{+46}_{-32}$ & 
$630^{+60}_{-60}$ TRGB; \citet{Ibata07} \\ 
&   &   &   &   &  $770^{+70}_{-70}$ TRGB; \citet{Letarte09} \\ [1.0ex]
\hline  
And XVI & $23.39^{+0.19}_{-0.14}$ & $0.066$  & $476^{+44}_{-29}$ & $319^{+43}_{-27}$ & $525^{+50}_{-50}$ TRGB; \citet{Ibata07} \\ 
&   &   &   &   &  $525^{+50}_{-50}$ TRGB; \citet{Letarte09} \\ [1.0ex]
\hline 
And XVII & $24.31^{+0.11}_{-0.08}$ & $0.075$  & $727^{+39}_{-25}$ & $67^{20}_{-24}$ & $794^{+40}_{-40}$ TRGB; \citet{Irwin08} \\  [1.0ex]
\hline
And XVIII & $25.42^{+0.07}_{-0.08}$ & $0.104$  & $1214^{+40}_{-43}$ & $457^{+39}_{-47}$ & $1355^{+88}_{-88}$ TRGB; \citet{McConn08} \\  [1.0ex]
\hline
And XIX & $24.57^{+0.08}_{-0.43}$ & $0.062$  & $821^{+32}_{-148}$ & $115^{+96}_{-9}$ & $933^{+61}_{-61}$ TRGB; \citet{McConn08} \\  [1.0ex]
\hline
And XX & $24.35^{+0.12}_{-0.16}$ & $0.058$  & $741^{+42}_{-52}$ & $128^{+28}_{-5}$ & $802^{+297}_{-96}$ TRGB; \citet{McConn08} \\  [1.0ex]
\hline
And XXI & $24.59^{+0.06}_{-0.07}$ & $0.093$  & $827^{+23}_{-25}$ & $135^{+8}_{-10}$ & $859^{+51}_{-51}$ TRGB; \citet{Martin09} \\  [1.0ex]
\hline
And XXII (Tri I) & $24.82^{+0.07}_{-0.36}$ & $0.075$  & $920^{+32}_{-139}$ & $275^{+8}_{-60}$ & $794^{+239}_{-0}$ TRGB; \citet{Martin09} \\  [1.0ex]
\hline
And XXIII & $24.37^{+0.09}_{-0.06}$ & $0.066$  & $748^{+31}_{-21}$ & $127^{+7}_{-4}$ & $733^{+23}_{-22}$ TRGB; \citet{Conn11} \\ 
&   &   &    &  &  $767^{+44}_{-44}$ HB; \citet{Richardson11} \\ [1.0ex]
\hline
And XXIV & $24.77^{+0.07}_{-0.10}$ & $0.083$  & $898^{+28}_{-42}$ & $169^{+29}_{-29}$ & $600^{+33}_{-33}$ HB; \citet{Richardson11} \\  [1.0ex]
\hline
And XXV & $24.33^{+0.07}_{-0.21}$ & $0.101$  & $736^{+23}_{-69}$ & $90^{+57}_{-10}$ & $812^{+46}_{-46}$ HB; \citet{Richardson11} \\  [1.0ex]
\hline
And XXVI & $24.39^{+0.55}_{-0.53}$ & $0.110$  & $754^{+218}_{-164}$ & $103^{+234}_{-3}$ & $762^{+42}_{-42}$ HB; \citet{Richardson11} \\  [1.0ex]
\hline
And XXVII & $25.49^{+0.07}_{-1.03}$ & $0.080$  & $1255^{+42}_{-474}$ & $482^{+0}_{-425}$ & $827^{+47}_{-47}$ HB; \citet{Richardson11} \\  [1.0ex]
\hline
And XXX$^\dagger$ (Cass II) & $24.17^{+0.10}_{-0.26}$ & $0.166$  & $681^{+32}_{-78}$ & $145^{+95}_{-4}$ & $565^{+25}_{-25}$ TRGB g-band; \citet{Irwin12} \\  [1.0ex]
\hline
NGC147 & $24.26^{+0.06}_{-0.06}$ & $0.173$  & $712^{+21}_{-19}$ & $118^{+15}_{-15}$ & $675^{+27}_{-27}$ TRGB; \citet{McConn05} \\  [1.0ex]
\hline
NGC185 & $23.96^{+0.07}_{-0.06}$ & $0.182$  & $620^{+19}_{-18}$ & $181^{+25}_{-20}$ & $616^{+26}_{-26}$ TRGB; \citet{McConn05} \\  [1.0ex]
\hline
M33 &  $24.57^{+0.05}_{-0.05}$ &  $0.042$  & $820^{+20}_{-19}$ & $210^{+6}_{-5}$ & $809^{+24}_{-24}$ TRGB; \citet{McConn05} \\ [1.0ex]
&   &   &    &  &  $964^{+54}_{-54}$ DEB; \citet{Bonanos06} \\ [1.0ex]
\hline

\end{tabular} 

\label{sat_table}  

\begin{flushleft} $Note:$ Extinction values are for object centers. Actual calculations apply individual corrections to each member star according to their coordinates. \end{flushleft}

\begin{flushleft} Distance derivation methods: TRGB = Tip of the Red Giant Branch; Ceph = Cepheid Period-Luminosity Relation; RR Ly = RR Lyrae Period-Luminosity Relation; RC = Red Clump; HB = Horizontal Branch; DEB = Detached Eclipsing Binary \end{flushleft}

\begin{flushleft} $^\dagger$Andromeda XXX is a new discovery, and will also be known as Cassiopeia II, being the second dwarf spheroidal satellite of M31 to be discovered in the constellation of Cassiopeia - see \citet{Irwin12}  \end{flushleft}

\end{table*} 




 
\begin{figure*}[htbp]
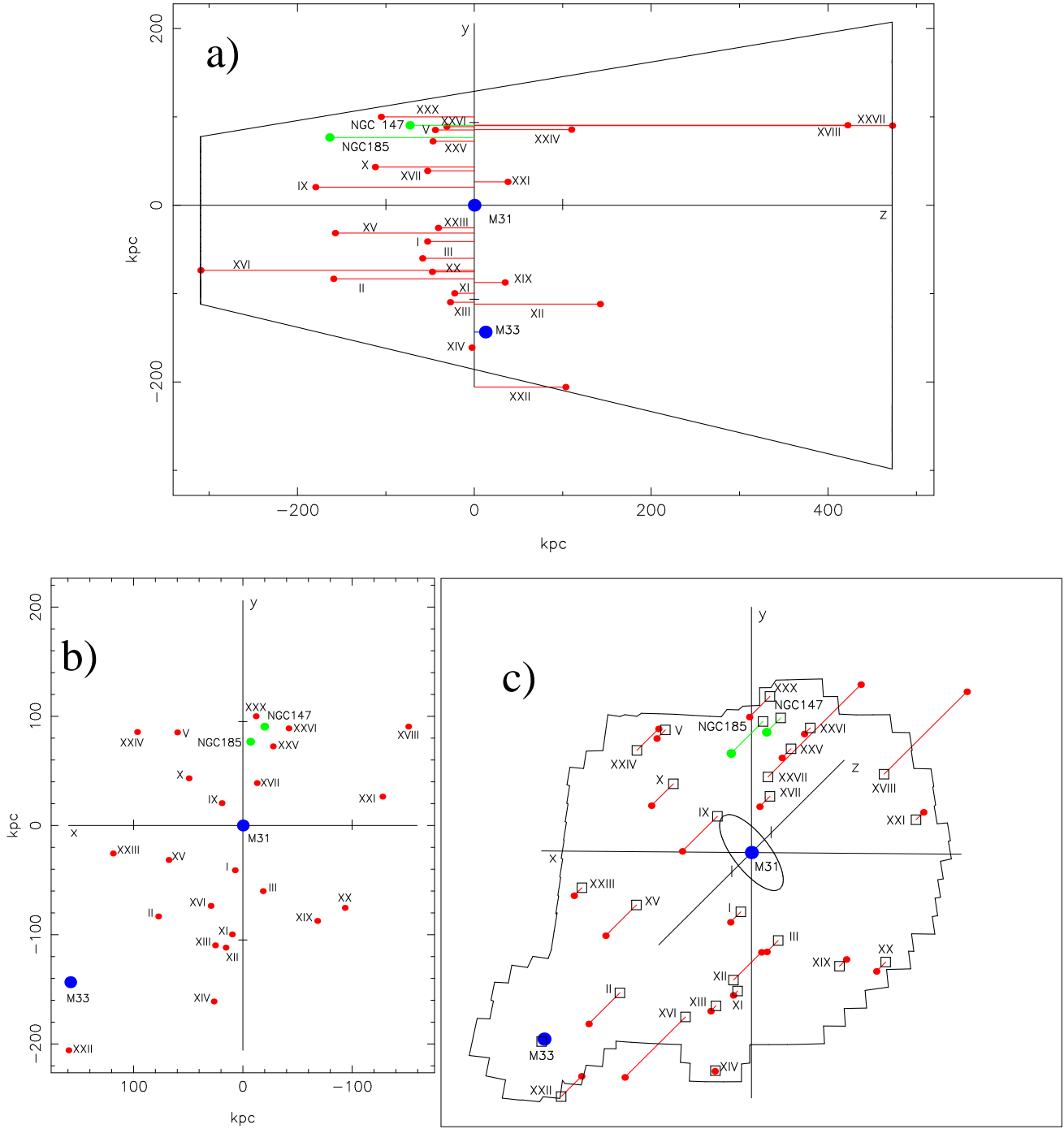

\begin{center}
		
\begin{overpic}[width = 0.5\textwidth,angle=-90]{Figures/M31_neighborhood_yz.ps}	
\put(10,60){\huge a)}
\end{overpic}

$ \begin{array}{cc} 
\\ \begin{overpic}[width = 0.5\textwidth,angle=-90]{Figures/M31_neighborhood_xy.ps}
\put(10,83){\huge b)}
\end{overpic}
\ \begin{overpic}[width = 0.5\textwidth,angle=-90]{Figures/M31_neighborhood_xyz.ps}
\put(10,70){\huge c)}
\end{overpic}

\end{array}$

\caption{Three views of the M31 Neighborhood: a) A view of the satellites of M31 along the y-z plane. The conic section illustrates the extent of volume covered by the PAndAS footprint as a function of distance from Earth; b) A view of the satellites of M31 in the x-y plane, revealing their true positions on the x-y plane after removing the effects of perspective (assuming the distances quoted in column 4 of Table \ref{sat_table}). Note that Andromeda XXVII lies directly behind NGC147 in this plot and is not labeled.; c) A 3D view of the satellites of M31. The satellite positions on the PAndAS footprint are indicated (i.e. with perspective conserved) along with the z-vector giving distance from the M31- centered tangent plane. The central ellipse indicates the approximate area of the survey where satellite detection is hindered by the M31 disk; Note: The perpendicular bars on relevant axes indicate 100 kpc intervals.}

\label{M31in3D}

\end{center}

\end{figure*}




\begin{figure*}[htbp]
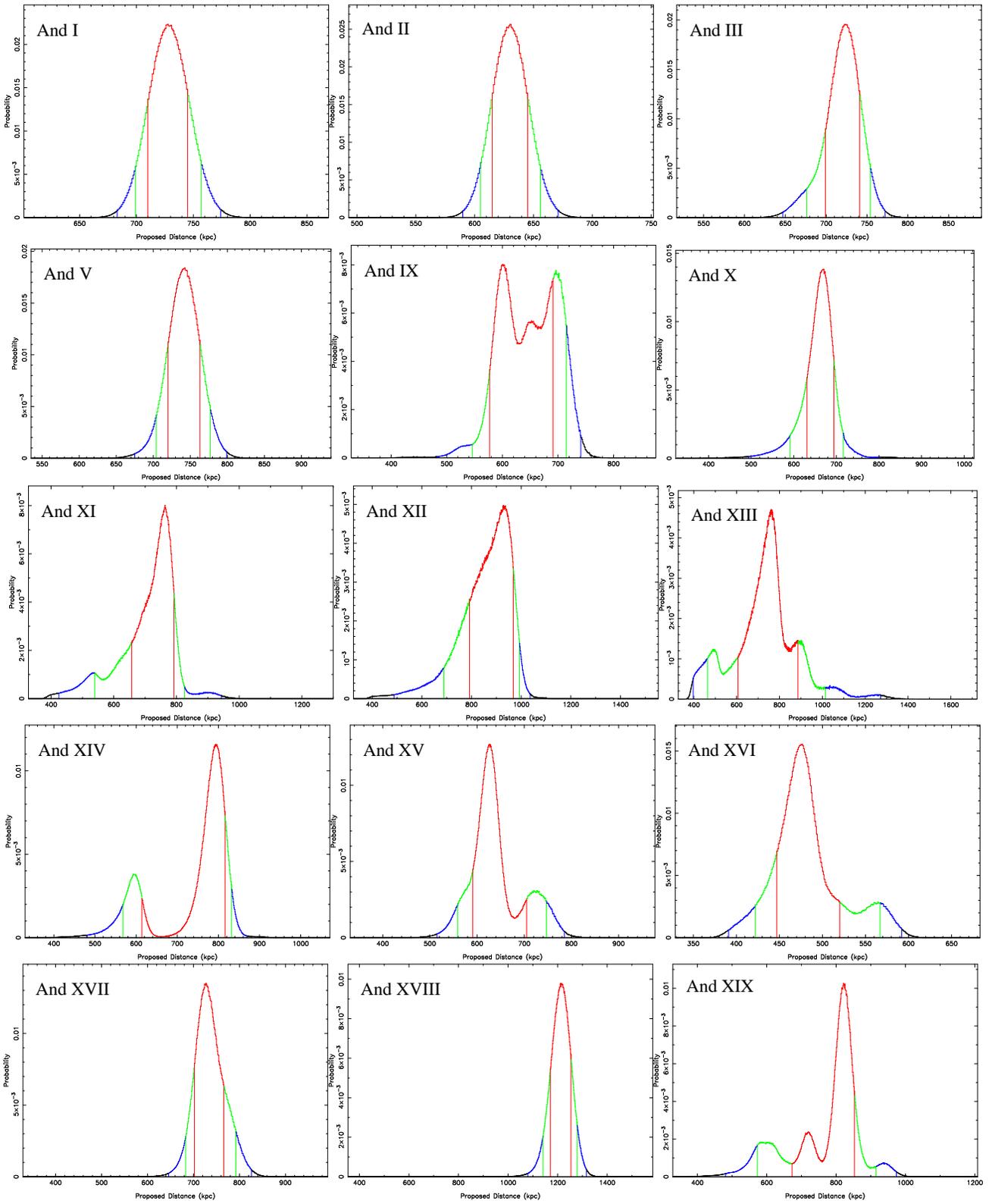

\begin{center} 
$ \begin{array}{c}
\begin{overpic}[width = 0.23\textwidth,angle=-90]{Figures/ANDI_dist_PPD.ps}
\put(10,63){\small And I}
\end{overpic}
\begin{overpic}[width = 0.23\textwidth,angle=-90]{Figures/ANDII_dist_PPD.ps}
\put(10,63){\small And II}
\end{overpic}
\begin{overpic}[width = 0.23\textwidth,angle=-90]{Figures/ANDIII_dist_PPD.ps}
\put(10,63){\small And III}
\end{overpic}
\\ \begin{overpic}[width = 0.23\textwidth,angle=-90]{Figures/ANDV_dist_PPD.ps}
\put(10,63){\small And V}
\end{overpic}
\begin{overpic}[width = 0.23\textwidth,angle=-90]{Figures/ANDIX_dist_PPD.ps}
\put(10,63){\small And IX}
\end{overpic}
\begin{overpic}[width = 0.23\textwidth,angle=-90]{Figures/ANDX_dist_PPD.ps}
\put(10,63){\small And X}
\end{overpic}
\\ \begin{overpic}[width = 0.23\textwidth,angle=-90]{Figures/ANDXI_dist_PPD.ps}
\put(10,63){\small And XI}
\end{overpic}
\begin{overpic}[width = 0.23\textwidth,angle=-90]{Figures/ANDXII_dist_PPD.ps}
\put(10,63){\small And XII}
\end{overpic}
\begin{overpic}[width = 0.23\textwidth,angle=-90]{Figures/ANDXIII_dist_PPD.ps}
\put(10,63){\small And XIII}
\end{overpic}
\\ \begin{overpic}[width = 0.23\textwidth,angle=-90]{Figures/ANDXIV_dist_PPD.ps}
\put(10,63){\small And XIV}
\end{overpic}
\begin{overpic}[width = 0.23\textwidth,angle=-90]{Figures/ANDXV_dist_PPD.ps}
\put(10,63){\small And XV}
\end{overpic}
\begin{overpic}[width = 0.23\textwidth,angle=-90]{Figures/ANDXVI_dist_PPD.ps}
\put(10,63){\small And XVI}
\end{overpic}
\\ \begin{overpic}[width = 0.23\textwidth,angle=-90]{Figures/ANDXVII_dist_PPD.ps}
\put(10,63){\small And XVII}
\end{overpic}
\begin{overpic}[width = 0.23\textwidth,angle=-90]{Figures/ANDXVIII_dist_PPD.ps}
\put(10,63){\small And XVIII}
\end{overpic}
\begin{overpic}[width = 0.23\textwidth,angle=-90]{Figures/ANDXIX_dist_PPD.ps}
\put(10,63){\small And XIX}
\end{overpic}

\end{array}$
\end{center}
\caption{Distance Posterior Distributions for dwarf spheroidal satellites And I - III, And V and And IX - XIX. The distributions are color-coded with red, green and blue denoting $1$-$\sigma$ ($68.2$\%), $90$ \% and $99$ \% credibility intervals respectively. The credibility intervals are measured from either side of the highest peak.}
\label{Dist_PPDs1}

\end{figure*}




\begin{figure*}[htbp]
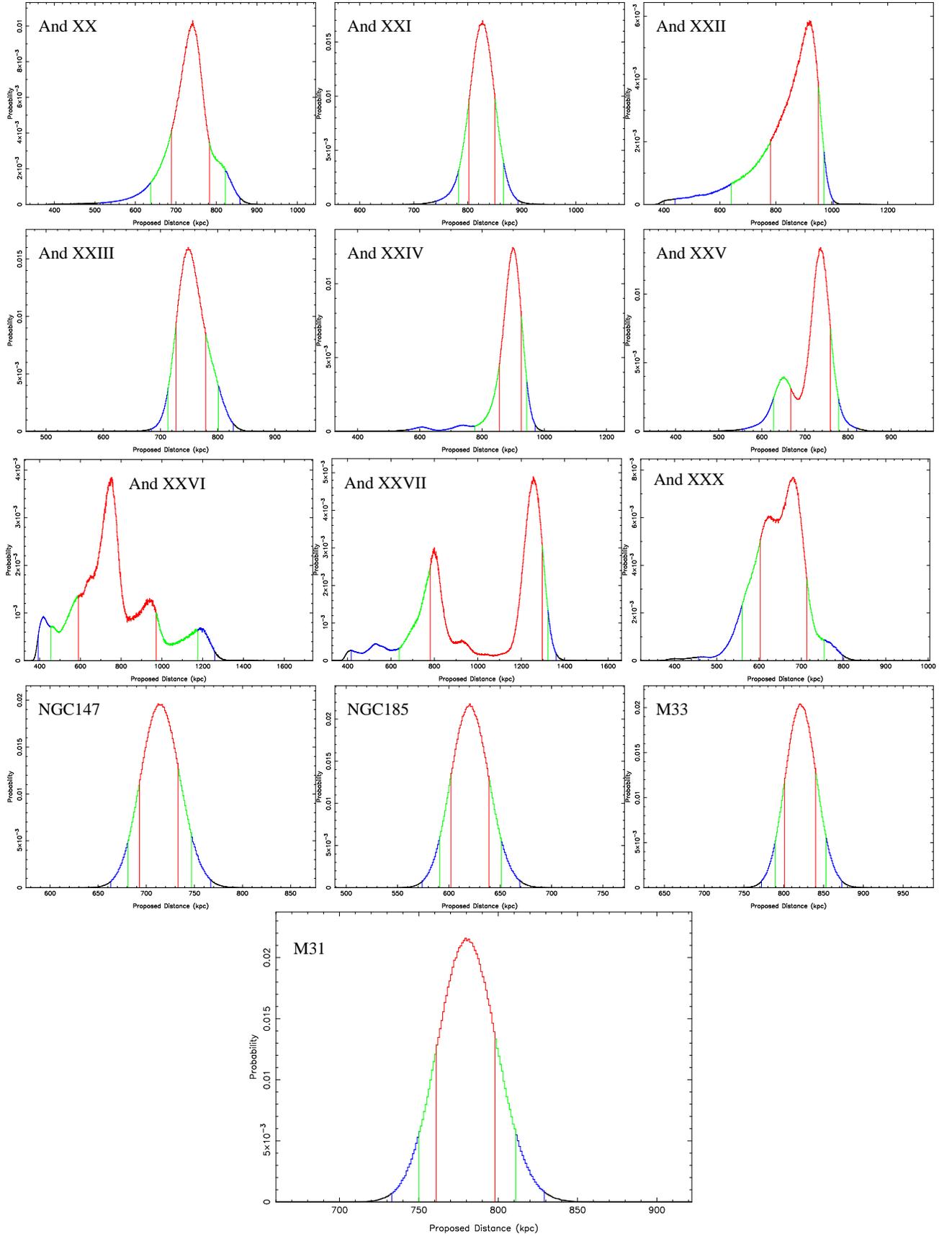

\begin{center} 
$ \begin{array}{c}
\begin{overpic}[width = 0.23\textwidth,angle=-90]{Figures/ANDXX_dist_PPD.ps}
\put(10,63){\small And XX}
\end{overpic}
\begin{overpic}[width = 0.23\textwidth,angle=-90]{Figures/ANDXXI_dist_PPD.ps}
\put(10,63){\small And XXI}
\end{overpic}
\begin{overpic}[width = 0.23\textwidth,angle=-90]{Figures/ANDXXII_dist_PPD.ps}
\put(10,63){\small And XXII}
\end{overpic}
\\ \begin{overpic}[width = 0.23\textwidth,angle=-90]{Figures/ANDXXIII_dist_PPD.ps}
\put(10,63){\small And XXIII}
\end{overpic}
\begin{overpic}[width = 0.23\textwidth,angle=-90]{Figures/ANDXXIV_dist_PPD.ps}
\put(10,63){\small And XXIV}
\end{overpic}
\begin{overpic}[width = 0.23\textwidth,angle=-90]{Figures/ANDXXV_dist_PPD.ps}
\put(10,63){\small And XXV}
\end{overpic}
\\ \begin{overpic}[width = 0.23\textwidth,angle=-90]{Figures/ANDXXVI_dist_PPD.ps}
\put(40,63){\small And XXVI}
\end{overpic}
\begin{overpic}[width = 0.23\textwidth,angle=-90]{Figures/ANDXXVII_dist_PPD.ps}
\put(10,63){\small And XXVII}
\end{overpic}
\begin{overpic}[width = 0.23\textwidth,angle=-90]{Figures/ANDXXX_dist_PPD.ps}
\put(10,63){\small And XXX}
\end{overpic}
\\ \begin{overpic}[width = 0.23\textwidth,angle=-90]{Figures/NGC147_dist_PPD.ps}
\put(10,63){\small NGC147}
\end{overpic}
\begin{overpic}[width = 0.23\textwidth,angle=-90]{Figures/NGC185_dist_PPD.ps}
\put(10,63){\small NGC185}
\end{overpic}
\begin{overpic}[width = 0.23\textwidth,angle=-90]{Figures/M33_dist_PPD.ps}
\put(10,63){\small M33}
\end{overpic}
\\ \begin{overpic}[width = 0.33\textwidth,angle=-90]{Figures/M31_dist_PPD.ps}
\put(10,63){\small M31}
\end{overpic}

\end{array}$
\end{center}
\caption{Distance Posterior Distributions for dwarf spheroidal satellites And XX - XXVII and And XXX, dwarf elliptical satellites NGC147 and NGC185, and major galaxies M31 and M33. The distributions are color-coded with red, green and blue denoting $1$-$\sigma$ ($68.2$\%), $90$ \% and $99$ \% credibility intervals respectively.}
\label{Dist_PPDs2}

\end{figure*}




\subsection{Determining the distances from M31}
\label{M31_Dist}

Once a satellite's distance from Earth is determined, it is straightforward to determine the distance from M31 using the cosine rule:

\begin{equation}
	r =  \sqrt{d^2 + (d_{M31})^2 - 2 d d_{M31} cos(\theta)}
	\label{e_CSR}
\end{equation}
where $r$ is the satellite's distance from M31, $d$ is the distance of the satellite from Earth, $d_{M31}$ is the distance of M31 from Earth and $\theta$ is the angle on the sky between M31 and the satellite. For convenience, we use a small angle approximation equating $\theta$ with its M31 tangent plane projection and note that any displacement of $r$ is insignificant due to the size of the $1 \sigma$ errors. If the uncertainty in distance to both M31 and the satellite takes on a Gaussian distribution, it is straightforward to determine the error in the satellite-M31 separation by adding the individual errors in quadrature. While it is reasonable to approximate the M31 distance uncertainty distribution as a Gaussian, the same cannot be said for each of the companion satellites. Hence once again it is more appropriate to sample values from the individual distance probability distributions. Thus, a histogram of $r$ values for the satellite is built up by sampling $d$ and $d_{M31}$ from their respective distributions over many iterations. This brings to the fore an important consideration: there is an integrable singularity in the resulting distribution at the closest approach distance to M31 ($r_c = d_{M31}sin(\theta)$) as shown below.    
	
The probability distribution for the Satellite-to-Earth distance $P(d)$ is related to that of the Satellite-to-M31 distance $P(r)$ as follows:
	  
\begin{equation}
	P(r) = \frac{\delta d}{\delta r} P(d)
	\label{e_Rdistn_to_Ddistn}
\end{equation}
From Eq. \ref{e_CSR}, and further noting that the Satellite-to-Earth distance corresponding to $r_c$ is $d_c = d_{M31}cos(\theta)$  we have:

\begin{equation}
	\frac{\delta d}{\delta r} = \frac{r}{d - d_c} 
\end{equation}
which allows us to derive:

\begin{equation}
	P(r) = \frac{r}{\sqrt{r^2 - r_c^2}} P(d)
\end{equation}
thus producing the singularity at $r = r_c$. In practice, after factoring in the Gaussian distribution in $d_{M31}$, this results in a sharp peak at the minimum possible Satellite-to-M31 distance when dealing with the more asymmetric Satellite-to-Earth distance probability distributions. Hence when considering the distribution of satellites as a function of distance from M31, one can either take the distances as determined directly from Eq. \ref{e_CSR} using solely the most likely distance from the Satellite-to-Earth distance distributions, or the whole distance probability distribution for a satellite can be allowed to influence the calculations, as accomplished via sampling. The final result can be quite different, depending on the choice.  


\subsection{A First Approximation of the Satellite Density Profile within the Halo} 
\label{ss_halo_analysis}

In the completed PAndAS Survey, we have for the first time a comprehensive coverage of a galaxy halo, with a \emph{uniform} photometric depth sufficient to identify even the comparatively faint satellite companions. In addition, in this paper we have provided distances to every one of these objects, all obtained via the \emph{same} method. We are thus presented with an excellent opportunity to study the density of satellites as a function of radius within a Milky-Way-like halo. 

As hinted at in the previous section, obtaining an accurate picture of the Satellite Density Profile (SDP) is not a trivial task. The first major consideration is to devise a way of factoring in the selection function. Comprehensive though the survey coverage is, it is not symmetric, and it is not infinite. Secondly, the choice of model for the SDP is not arbitrary. Whether a simple, unbroken power law is sufficient is not immediately clear. Furthermore, does it even make any sense to treat the halo as a radially symmetric, isotropic distribution? A glance at the obvious asymmetry in Fig. \ref{M31in3D} (a) would suggest otherwise. Nevertheless, for a first approximation it is reasonable to consider what the best fitting radially-symmetric, unbroken power law to the SDP would be. 

The PAndAS Survey covers approximately 400 square degrees of sky and is roughly symmetric about the center of the M31 disk but with a major protrusion in the South East to encompass the M33 environs. For the purpose of obtaining an accurate measure of the survey coverage of the halo as a function of radius, as well as factoring in the actual survey borders, an inner ellipse was also subtracted where the presence of the M31 disk has made satellite detection more difficult. Both the outer survey borders and the inner cut-off ellipse are plotted in Fig. \ref{M31in3D}. The inner cut-off ellipse has an eccentricity $\epsilon = \sqrt{0.84}$ and is inclined with the semi-major axis angled $51.9^{\circ}$ with respect to the $x$-axis $(\eta = 0)$. The dwarf galaxies M32 and M110 lie inside this ellipse as do the somewhat dubious satellite identifications Andromeda VIII and Andromeda IV (see \citealt{Ferguson2000}), hence their omission from the data presented in Table \ref{sat_table}. With the inner and outer boundaries suitably delineated, the procedure then was to determine what fraction of halo volume at a given radius $f(r)$ would fall within these boundaries once projected onto the M31 tangent plane. This was achieved by implementing the Even-Odd Rule on the projections of uniformly populated halo shells.      

Having determined $f(r)$, we can proceed to determine the required normalization for a power law of any given $\alpha$, allowing us to use the power law directly as a probability distribution. Setting the problem out in terms of probabilities, we require to determine the probability of each tested M31-to-object distance (henceforth simply `radius') $r$ given a power law with slope $\alpha$:

\begin{equation}
	P(r|\alpha) = \frac{k}{r^{\alpha}}
\end{equation}
where $k$ is the normalization constant and $r_{min} \le r \le r_{max}$. Using the assumed spherical symmetry, we then have:

\begin{equation}
	f(r) \int_{r_{min}}^{r_{max}} P(r|\alpha) \int_{0}^{2\pi} \int_{0}^{\pi} r^2 sin \theta d\theta d\phi dr = 1
\end{equation}
so that

\begin{equation}
	4 \pi f(r) \int_{r_{min}}^{r_{max}} kr^{2 - \alpha} dr = 1
\end{equation}
Hence, for a given radius at a given $\alpha$, we have:

\begin{equation}
	k(r, \alpha) = \left[4 \pi f(r) \left( \frac{r^{3-\alpha}}{3-\alpha} \right)_{r_{min}}^{r_{max}}\right]^{-1}.
\end{equation}

The calculation of the likelihood for a power law of a given slope $\alpha$ may be simplified by noting that for any given radius, $f(r)$ and hence $k$ act to scale the probability in an identical way whatever the value of $\alpha$. Thus the dependance of $k$ on $r$ is effectively marginalized over when the posterior distribution for $\alpha$ is calculated, so long as any sampling of radii utilizes the same radii at every value of $\alpha$. The likelihood for a given power law (i.e. a given $\alpha$) is thus:

\begin{equation}
	{\cal L(\alpha)} = \prod_{n=i}^{nsat} k r_i^{2 - \alpha}
\end{equation}
where $nsat$ is the number of satellites - i.e. the 27 companions of M31 listed in Table \ref{sat_table}. As discussed in \S \ref{M31_Dist}, there are essentially two ways we can determine the likelihood of a given $\alpha$. The most straightforward is to use single values of $r_i$ as determined directly from the mode in the posterior distribution for each satellite using Eq. \ref{e_CSR}. The second, and arguably more robust method is to use the entire radius probability distribution (RPD) for each satellite. In the case of this second approach, the likelihood for the power law determined for each satellite becomes a convolution of the power law with the satellite's RPD, so that the likelihoods of the individual samples are summed. The final likelihoods determined for each satellite can then be simply multiplied as before, giving a total likelihood as follows:

\begin{equation}
\begin{split}
{\cal L(\alpha)} &= (k r_{1,1}^{2-\alpha} + k r_{2,1}^{2-\alpha} + \ldots + k r_{nsam,1}^{2-\alpha}) \; \times \\
	    & \quad(k r_{1,2}^{2-\alpha} + k r_{2,2}^{2-\alpha} + \ldots + k r_{nsam,2}^{2-\alpha}) \times \ldots \times\\
	    & \quad(k r_{1,nsat}^{2-\alpha} + k r_{2,nsat}^{2-\alpha} + \ldots + k r_{nsam,nsat}^{2-\alpha}) \\
	    & \bigskip \\
		&= \prod_{n=i}^{nsat} \left[ \sum_{n=j}^{nsam} k r_{j,i}^{2-\alpha}\right]
\end{split}
\end{equation}
where $r_{j,i}$ is the $j^{th}$ sampled radius of the $i^{th}$ satellite, and $nsam$ is the total number of samples. 

  
\begin{figure}[htbp]
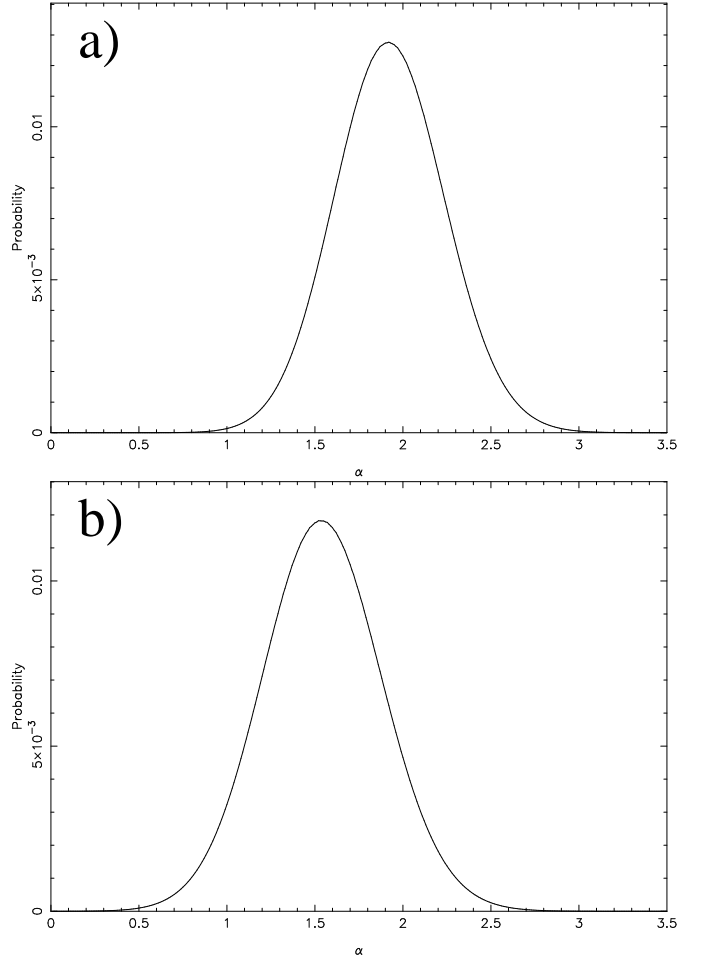

\begin{center}

$ \begin{array}{c} 
\begin{overpic}[width = 0.35\textwidth,angle=-90]{Figures/ns_alpha_PPD.ps}
\put(10,63){\huge a)}
\end{overpic}
\\ \begin{overpic}[width = 0.35\textwidth,angle=-90]{Figures/500000samp_alpha_PPD.ps}
\put(10,63){\huge b)}
\end{overpic}

\end{array}$

\caption{Probability distributions for the slope $\alpha$ of a single power law used to model the M31 halo satellite distribution, given the entire set of 27 M31 companions presented in Table \ref{sat_table}. Figure (a) gives the distribution assuming a single best fit radius for each of the satellites as determined from the mode in the satellite's distance posterior distribution (as given in Column 4 of Table \ref{sat_table}). Figure (b) shows the same distribution when the entire radius probability distribution for each satellite is sampled 500000 times.}

\label{fig_halodensity}

\end{center}

\end{figure}


The resulting distribution achieved by implementing the first approach is presented in Fig. \ref{fig_halodensity} (a), from which a value for $\alpha$ of $1.92_{-0.30}^{+0.32}$ is obtained. It is interesting to note that this value is consistent with an isothermal satellite distribution with uniform velocity dispersion. Replacing the individual best-fit radii with 500000 samples from the respective RPD for each satellite as per the second approach, the result is substantially different, as demonstrated by Fig. \ref{fig_halodensity} (b). Here a value for $\alpha$ of $1.52_{-0.32}^{+0.35}$ provides the best fit to the data. This discrepancy is presumably a consequence of the non-Gaussian RPD profiles for the more poorly populated satellites, as noted in \S \ref{M31_Dist}. In fact, if the 15 most Gaussian-like distributions are taken alone, namely Andromedas I, II, III, V, X, XVI, XVII, XVIII, XX, XXI, XXIII, XXIV, NGC147, NGC185 and M33, the results are in much closer agreement, with $\alpha = 1.87_{-0.42}^{+0.46}$ with sampling and $\alpha = 2.02_{-0.41}^{+0.43}$ without. 

Given the obvious asymmetry in the satellite distribution in Fig. \ref{M31in3D}, it is interesting to consider the effects of isolating various other satellites from the calculations. The stark asymmetry between the number of satellites on the near side as opposed to the far side of the M31 tangent plane for instance (as had been initially reported by \citealt{McConn06}) is echoed in the respective density profiles, with an $\alpha$ of $2.37_{-0.37}^{+0.42} \; (no sampling)$ recorded when only the near-side satellites are considered, and that of $0.93_{-0.49}^{+0.56} \; (no sampling)$ when instead the far-side galaxies alone are included. When the individual satellite RPDs are sampled, the corresponding values are $1.87_{-0.40}^{+0.43}$ and $0.78_{-0.46}^{+0.61}$ respectively. Despite the large uncertainties, the results clearly do not support symmetry of any kind about the tangent plane. It is important to note however, that this asymmetry may not be physical, but rather an effect of incompleteness in the data at the fainter magnitudes of the satellites on the far side of M31. \citet{McConn06} do however observe this asymmetry even when only the more luminous satellites are considered. In time, it is hoped that the nature of the data incompleteness will be better understood and effort \emph{is} underway to determine the completeness functions for dwarf galaxy detection in the PAndAS survey \citep{Martin12B}. In the mean time, it would seem prudent to regard the contribution to the density profile of the far side satellites with caution, instead taking the density profile measured from the near-side satellites alone as the best measurement. 

On a final note with regard to near-side-far-side asymmetry, it is important to realize that the uncertainty in the distance to M31 has a large effect on how many satellites will lie on either side of the M31 tangent plane, and indeed on the density measurement as a whole. Where the individual PPDs are sampled, this is taken into account as the M31 PPD is sampled for each measurement. Nevertheless, it is interesting to consider the specific (non-sampled) case where M31 is measured at a closer distance, while all best-fit satellite distances remain unchanged. From the M31 PPD in Fig. \ref{Dist_PPDs2} , it can be seen that there is a 5\% chance that M31 lies at 750 kpc or closer. If M31 is taken to lie at 750 kpc, Andromedas XI, XIII and XIV move onto the far side of the M31 tangent plane, going someway to even out the asymmetry. However, if the distances of all the satellites from M31 are re-measured for this new M31 position, the same stark contrast between the density profiles for the near and far sides remains and in fact grows. Using only those satellites on the near side of the new M31 tangent plane, an $\alpha$ of $2.87_{-0.45}^{+0.50}$ is determined whereas if only those satellites on the far side are considered, an $\alpha$ of $1.22_{-0.47}^{+0.47}$ is obtained. Hence it would seem unlikely that the observed near-side-far-side asymmetry is primarily a consequence of an overestimated M31 distance.

Recent research, such as that presented by \citet{Koch06} and \citet{Metz07} point toward highly significant planar alignments of various collections of satellites within the M31 halo, even though as a whole, no such distribution is prominent. Interestingly, the former investigation finds that it is predominantly the objects morphologically similar to the dwarf spheroidals in their sample that can be constrained to a relatively thin disk, which also includes NGC147 and M33. While our sample is considerably larger, it nevertheless consists nearly entirely of such objects, so it will be interesting to determine what degree of symmetry may be found within and on either side of the best-fit plane. We intend to investigate this in an upcoming publication, though it must still be noted that outliers from the planar trend have already been noted in this small sample, such as Andromeda II and NGC185. Furthermore, other members are known not to conform to the norm of M31 satellite dynamics, with Andromeda XIV for instance apparently at the escape velocity for the M31 system for its determined distance \citep{Majewski07}. Indeed, it would seem that whatever model is assumed, a few outliers are inevitable.



\section{Conclusions}
\label{s_Conclusions} 

With the ready applicability of the TRGB standard candle to almost any of our galactic neighbors, there can be no question that its role will continue to be an important one. As the world's premiere telescopes grow in size, so too will the radius of the `neighborhood' of galaxies to which the TRGB can be applied. Hence a technique which accurately characterizes the true probability space of the TRGB distances determined is a great asset. Indeed this quality comes to play an increasingly important role as more and more sparsely populated objects are found to frequent the environs of our larger nearby neighbors. The differences in the results achieved in the previous section with and without sampling of the actual distance distributions illustrates this fact. 

Where in Paper I, the foundations were laid for a TRGB method with such desirable qualities, its full value only becomes apparent when one actually employs its full Bayesian potential. It only requires a brief glance at figures \ref{ANDX_LF_fits} and \ref{ANDX_PPDs} to see how powerful a single data-specific prior can be. Similarly, the simple distance weighting prior outlined in \S \ref{ss_add_prior} can make a poorly constrained model quite workable, as illustrated in the case of Andromeda XIII. Both tools will likely prove very useful when the method is used further afield. 

It should also be remembered that the TRGB standard candle is in many ways, just the `first assault.' When photometric data of sufficient depth is obtained, the horizontal branch can often pin down the distance with still greater accuracy. With a simple adjustment to the model luminosity function, the techniques outlined in this paper and its predecessor commute quite readily to implementation on the horizontal branch.

Lastly, It must also be said that the distances presented herein provide an excellent opportunity to provide a new, updated analysis of the asymmetry and density of the M31 halo satellite distribution, one only touched on here. With such comprehensive and consistent coverage, there is great potential in these distances to further constrain the possible evolution and dynamical history of the M31 halo system.


\begin{acknowledgments}
A. R. C. would like to thank Macquarie University for their financial support through the Macquarie University Research Excellence Scholarship (MQRES) and both the University of Sydney and Universite« de Strasbourg for the use of computational and other facilities.
R. A. I. and D. V. G. gratefully acknowledge support from the Agence Nationale de la Recherche though the grant POMMME (ANR 09-BLAN-0228). G. F. L. thanks the Australian Research Council for support through his Future Fellowship (FT100100268) and Discovery Project (DP110100678). M. A. F. acknowledges support from NSF grant AST-1009652. The analysis contained in this publication uses photometric data obtained with MegaPrime/MegaCam, a joint project of CFHT and CEA/DAPNIA, at the Canada-France-Hawaii Telescope (CFHT) which is operated by the National Research Council (NRC) of Canada, the Institute National des Sciences de l'Univers of the Centre National de la Recherche Scientifique of France, and the University of Hawaii. Our thanks go to the entire staff at CFHT for their great efforts and continuing support throughout the PAndAS project.
\end{acknowledgments}


\end{document}